\documentclass[printer]{aa} % for a referee version
\usepackage{graphicx}
\usepackage{txfonts}
\usepackage{natbib}
\usepackage{color}
\usepackage{hyperref}

\begin{document}

\title{Improved precision of radial velocity measurements after correction for telluric absorption}

%\subtitle{Improvement of high-precision radial velocities measurements by means of telluric absorption removal.}
%\shorttitle{telluric correction}

\author{A. Ivanova\inst{1,2}
\and
R. Lallement\inst{3}
\and
J.-L. Bertaux\inst{1,4}
}

\institute{LATMOS, Universit\'e Versailles-Saint-Quentin, 11 Bd D'Alembert, 78280 Guyancourt, France\\
              \email{anastasia.ivanova@latmos.ipsl.fr}
  \and
 Space Research Institute (IKI), Russian Academy of Science, Moscow 117997, Russia 
  \and
  GEPI, Observatoire de Paris, PSL University, CNRS,  5 Place Jules Janssen, 92190 Meudon, France
\and
LATMOS, Sorbonne Universit\'e, 4 Place Jussieu, 75005 Paris, France
}

\date{Received ; accepted }
\titlerunning{Telluric correction}
% \abstract{}{}{}{}{} 
% 5 {} token are mandatory
 
\abstract
% context heading (optional)
%leave it empty if necessary  
{The detection of planets around other stars by the measurement of the stellar Radial Velocity (RV) variations benefits from improvements of dedicated spectrographs, allowing to achieve a precision of 1 ms$^{-1}$ or better. Spectral intervals within which stellar lines are contaminated by telluric lines are classically excluded from the RV processing.}
% aims heading (mandatory)
{We aim at estimating the potential improvement of telluric absorption removal and subsequent extension of the useful spectral domain on the precision of radial velocity measurements.}
% methods heading (mandatory)
{We developed a correction method based on the on-line web service TAPAS, allowing to determine a synthetic atmospheric transmission spectrum for the time and location of observations. This method was applied to the telluric H$_{2}$O and O$_2$ absorption removal from a series of 200 ESPRESSO consecutive exposures of the K2.5V star HD40307, available in ESO archives. We calculated the radial velocity using the standard Cross-Correlation Function (CCF) method and Gaussian fit of the CCF, with uncorrected spectra and the ESPRESSO standard stellar binary mask on one hand, and telluric-corrected spectra and an augmented binary mask with 696 additional lines on the other hand.}
 % results
{We find that the precision of radial velocity measurements is improved in the second case, with a reduction of the average formal error from 1.04 ms$^{-1}$ down to 0.78 ms$^{-1}$ in the case of these ESPRESSO data and this stellar type for the red arm. Using an estimator of the minimal error based on photon noise limit applied to the full CCF, the error is reduced from 0.89 ms$^{-1}$ down to 0.78 ms$^{-1}$.  This corresponds to a significant decrease of about 35\% of observing time to reach the same precision in the red part.
}
{}

\keywords{methods:data analysis/ techniques:spectroscopic/ techniques:radial velocities/stars: planetary systems}

\maketitle

\section{Introduction}

After the discovery of the first exoplanet \citep{Mayor1995}, the radial velocity (RV) method has brought us another $\simeq$ 2000 exoplanets\footnote{https://exoplanetarchive.ipac.caltech.edu}. Although the method of transit photometry is the most productive at the moment, the RV method is still important because, in addition to being able to discover new exoplanets and determine their apparent mass m sin{\it i}, it is used to confirm exoplanets found by other methods and measure the mass of transit exoplanets (sin{\it i} $\simeq$1). However, the RV method is limited by technical capabilities: the size and availability of the telescope, the stability and accuracy of the wavelength calibration of the spectrograph. Therefore, any improvement of the processing associated to the radial velocity retrieval will result in a better achieved precision, or in a shorter telescope time required to achieve a defined-goal precision (when it is limited by photon shot noise).\\
The RV method \citep{Connes1985,Baranne1996} is based on measuring the Doppler shifts of the stellar lines, which is caused by the reflex motion of the star around the center of mass of the star-planet system. Due to the fact that spectra for RV measurements are taken from Earth, the absorption lines produced by Earth’s atmosphere are also implanted on the star spectrum, along with the stellar lines. Earth’s atmosphere absorptions, or telluric absorptions, are unavoidable. They produce a system of absorption features which wavelengths are essentially fixed in the wavelength reference system of the spectrograph. On the contrary, even if the radial velocity of a star w.r.t. the barycenter of the solar system is constant (case of no planet), the motion of the Earth (rotation and orbital) induces a change of Doppler shift of the star as seen from the telescope (Barycentric Earth Radial Velocity, BERV).\\
The classical method of measuring RV in spectra is to cross-correlate them with a template (binary mask, BM) \citep{Baranne1996}. At the very beginning binary masks were mechanical \citep{Baranne1979}, with holes at the positions of stellar absorption lines and refocusing of the whole spectrum light getting through the mask on a single photometric detector. Later on, masks were numerical, under the form of a set of boxcar functions \citep[e.g.][]{Baranne1996}. Subsequent improvements to the Cross Correlation Function (CCF) method consisted in assigning different weights to different lines of the BM \citep{Pepe2002,Lafarga2020}.

\begin{figure*}
\centering
\includegraphics[width=0.9\hsize]{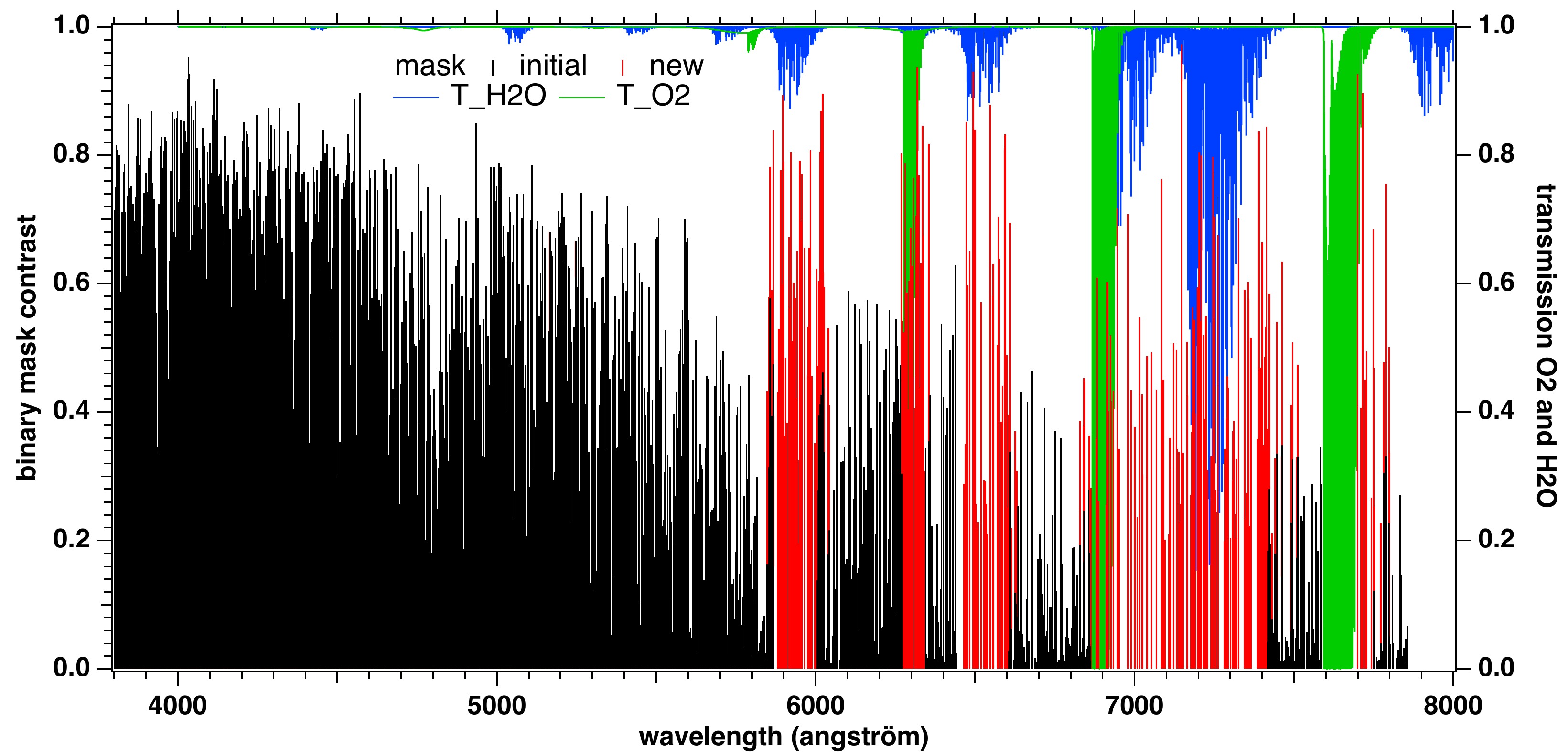}
\caption{Binary masks (left scale) and telluric lines (right scale). Black bars represent the contrast (relative depths) of the stellar lines of the standard K2 binary mask BM1 used in the ESPRESSO pipeline. Red bars show the contrast of the newly added lines forming the mask BMc.  TAPAS telluric H$_2$O  and O$_2$  transmission spectra are shown in blue and green respectively.}
\label{figure_1_mask}
\end{figure*}

\begin{figure}
\centering
\includegraphics[width=0.95\hsize]{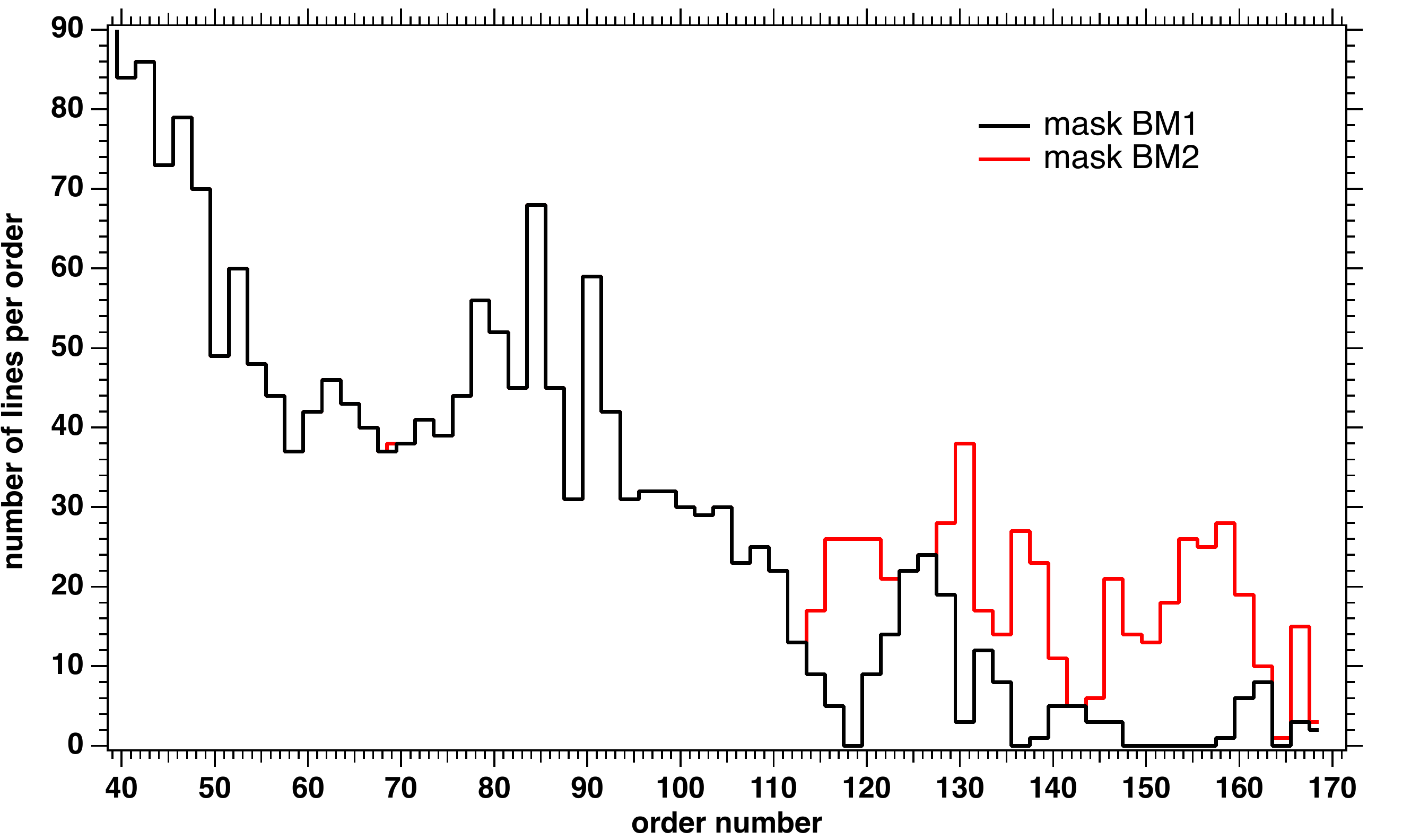}
\caption{Number of stellar lines per order from the standard BM1 (black) and the augmented binary mask BM2=BM1+BMc  (red) , restricted to lines with a contrast $\geq$0.2 and used in our study. For the red arm, the new mask contains 724 more lines than the 1060 lines of the original mask (Table \ref{Tab_numberoflines}).}
\label{figure_2_number_lines}
\end{figure}

Most RV measurements excluded from consideration spectral regions with telluric absorption or close to them. Fig.\ref{figure_1_mask} illustrates those wavelength intervals which are excluded from computation due to telluric contamination by restricting the official ESPRESSO BM lines to spectral regions clear of strong to moderate telluric lines. If we convert these regions into orders of the new, high accuracy ESPRESSO spectrograph at ESO-VLT, 16 orders out of 170 are not used in RV calculation due to this reason, while 30 others have a quite small number of lines of the binary mask (Figure \ref{figure_2_number_lines}). Our goal is to show that it is possible to correct the spectra from telluric absorption, based on a highly detailed state-of-the-art synthetic atmospheric transmission spectrum available on-line and a rather simple fitting technique, and to use the previously discarded spectral regions for RV measurement. To check the correction and test its benefits, we took advantage from a quite unusual series of very short ESPRESSO exposures made in a single night.

Several types of methods have been developed to correct spectra for telluric absorption. We describe types and differences in existing techniques, and also show the place of our method in this typology in section \ref{section_overview}. In section \ref{data}, we describe the archive ESPRESSO data that we have used for this study. Section \ref{telluric} describes the steps towards the removal of the telluric contamination in all exposures. In section \ref{RV} we recall the classical CCF method, and explain how we constructed a new binary mask relevant to corrected parts of the spectrum.  In Section \ref{results} are described the results when the CCF technique is applied to various parts of the spectrum, and the improvement in RV retrieval precision is quantified. In Appendix A we describe the method used to estimate the uncertainties on RV as a function of photon noise and spectrum quality factor, and how uncertainties derived in this way compare with our results. Finally, a comparison of radial velocity fluctuations seen with the blue arm and the red arm of ESPRESSO is conducted in section \ref{comparison}, followed by conclusions.

\section{Overview of existing techniques.}
\label{section_overview}
A first technique is the division by the spectrum of a fast-rotating star observed (the so-called telluric standard) as close as possible in time and direction to the science target. It has the disadvantage of requiring consequent additional observing time and the spectrum is affected by features present in the standard star. It is well adapted to multi-object spectrographs; in this case a number of fibers are devoted to the telluric target stars, but it is not well adapted to high precision radial velocity measurements. A second, data-driven category builds on series of observations.  Among these, \cite{Artigau14} used series of hot, fast rotating stars to construct libraries of empirical telluric absorption spectra and perform a Principal Component Analysis (PCA) of the dataset. The authors showed that the simultaneous adjustments of the first five principal components to an observed star spectrum allow to obtain a good correction and to extend the spectral domain for RV extraction and to increase the RV accuracy. More recently, \cite{Leet19} used series of fast rotator spectra observed in various different humidity conditions to empirically extract the whole series of water vapor lines, including the very weak, microtelluric absorptions. \cite{Bedell19} developed a modern, sophisticated method of learning template spectra for stars and tellurics directly from the data, a method adapted to series of observations of cool stars covering wide enough Doppler shifts between star and Earth. Recently, \cite{Cretignier21} developed a PCA on spectral time series with prior information to disentangle contaminations from real Doppler shifts, i.e. to eliminate both instrumental systematics and atmospheric contamination.

A third category of telluric decontamination relies on realistic atmospheric transmission spectra.  Extremely detailed models of the global atmosphere and of time- and location-dependent atmospheric profiles are now available, the most detailed being the Global Data Assimilation System (GDAS) and the European Centre for Medium-Range Weather Forecasts (ECMWF).  They give the altitude distribution of pressure, temperature, and H$_2$O concentration. Several tools have been constructed over the last decades to compute the resulting transmission spectra based on these profiles and the HITRAN molecular database \citep{Rothman2009, Gordon2017}. The MolecFit facility developed by \cite{Smette2015} and the Telfit tool by \cite{Gullikson2014} make use of GDAS atmospheric profiles, while the TAPAS online service \citep[see ][]{Bertaux2014} is extracting vertical profiles from the ECMWF. Both MolecFit and TAPAS use the state-of-the-art Line-By-Line Radiative Transfer Model (LBLRTM) code and the HITRAN database to compute the atmospheric transmission. Telfit  and MolecFit also include software allowing to correct spectral observations based on fits of the atmospheric transmission to the data, by adjusting the wavelength solution, a polynomial continuum and molecule abundances. To do so, Molecfit retrieves automatically the atmospheric profile at the time and place of the observations from an ESO repository. The advantage of this third category is the absence of needed additional measurements and the adaptability to any instrument and observatory.

Very recently \cite{Allart22} (hereafter A22) presented a fourth type of method to correct stellar spectra from telluric absorptions. Similarly to the previous techniques A22 use the HITRAN spectral data base to compute the transmittance of H$_{2}$O and O$_{2}$, however, instead of a worldwide time-variable grid of detailed atmospheric altitude profiles they model the  atmosphere as a single layer, characterized by a unique pressure p, unique temperature T and a global H$_{2}$O column. The line profiles are less accurate by essence, because the resulting atmospheric transmission does not take into account the different widths that correspond to different altitudes, contrary to the realistic atmosphere,  but the advantage of the technique is the absence of dependence on a meteorological field. A22 adjust p, T, and H$_{2}$O abundance by piling up 20 H$_{2}$O lines in a CCF.

Our work enters in the third category. Here, for the Earth’s atmospheric model and the simulated transmission spectra, we used the TAPAS facility \citep{Bertaux2014}. TAPAS\footnote{https://tapas.aeris-data.fr/en/home/} is a free online web service that simulates with exquisite details the atmospheric transmission spectra due to the main absorbing species as a function of date, observing site and either zenith angle or target coordinates.  It is using the ESPRI/AERIS ETHER (a French Atmospheric Chemistry Data Center) facility to obtain pressure, temperature and constituent vertical profiles of the ECMWF, refreshed every 6 hours. These are used to compute the atmospheric transmittance from the top of the atmosphere down to the observatory in each of 137 altitude layers, based on the HITRAN database and LBLRTM  \citep{Clough1995, Clough2005}. For each species and each altitude layer the local temperature and pressure are used to compute air- and self-broadening as well as pressure line shift for each HITRAN transition. The basic mode gives access to spectra with  wavelength grid steps on the order of 1 m\AA\ (R $\simeq$ 5 10$^{6}$), and the transmission spectra may be obtained separately for individual gases. TAPAS does not include correction tools and serves as a basis for line identifications or for various post-processing correction techniques.  For example, \cite{Artigau21} used TAPAS for a first data/model adjustment followed by a refined PCA-based correction for residuals. \cite{Bertaux2014} described the TAPAS product, a synthetic atmospheric transmission spectrum for the time, place and direction of any astronomical observation. The potential application to the RV method of exo-planets search was mentioned, and the way to adapt the H$_2$O quantity of the model to a real observation was only sketched, with the use of T$^{X}$(H$_2$O), which is the model transmission T(H$_2$O) elevated to the power X, to be determined by comparison with the observed spectrum. In the present paper, we put in practice the method sketched in this paper.

\section{Data}
\label{data}
\subsection{Brief description of the ESPRESSO spectrograph}
The new-generation spectrograph ESPRESSO was chosen because of its high resolution, its excellent wavelength calibration, and especially its impressive stability and quality of data.
Indeed, ESPRESSO can achieve a precision of measurement of 30 cms$^{-1}$, as reported in \cite{Pepe2021}.
The authors present a detailed description of the spectrograph, as well as an assessment of performances. 
Installed by ESO at Paranal observatory, the ESPRESSO spectrograph can be coupled to any of the four VLT telescopes by optical fibers. It is an echelle-grating spectrograph with cross-dispersion, installed in a vacuum chamber. 
It covers the range 378.2 to 788.7 nm wavelength range, divided in two arms with two different detectors. We have used the High Resolution mode, providing a resolving power somewhat variable with position on the detectors, between $\simeq$100000 and $\simeq$160000. In this mode, an image slicer is used, providing two images of the same echelle-grating physical diffraction order. All the useful parts of spectra are numbered from 0 (start of the blue arm) to 169 (end of the red arm), and are called «orders» in the official nomenclature of ESPRESSO (which will be used throughout this paper), not to be confused with true echelle-grating diffraction orders. In this nomenclature, one even number order is followed by an odd order which has the same wavelength coverage (due to the image slicer). 
The ESPRESSO processing pipe-line has several versions of outputs. We selected the ES\_S2BA-type data, i.e. one spectrum per order, without flux calibration, not corrected for blaze function, with 9111 spectels in the red and  6541 in the blue. We selected the wavelength assignment of each spectel in the laboratory system, in which the telluric lines absorption system should be fixed (except for the local wind). The blue arm contains orders 0 to 89 and covers the range 3781-5251 \AA~, while the red arm contains orders 90 to 169 and covers 5195-7908\AA~. 
The intensity for each spectel is given in adu (analog to digital units). The electronic gain that was used for the archive data discussed here is $\simeq$0.9 electron/adu.

\subsection{Selected data}

We have used ESO VLT/ESPRESSO spectra of the K2.5V star HD40307 (V=7.1). The star has a mean radial velocity of $\simeq$ 31.3 km$s^{-1}$ and at least four discovered exoplanets with periods from 4.3 to 197 days \citep{Diaz2016}. The target star HD40307 was chosen for two main reasons: first, detailed measurements of radial velocities of this target with ESPRESSO were obtained as part of an astero-seismological campaign on the star\footnote{Prog.ID:0102.D-0346(A); PI: Bouchy}. The results of the full campaign are contained in  \cite{Pepe2021}.\\ Second, the ESO archive makes available all 1150 spectra of this star recorded over 5 nights. Here we made use of the set of 200 consecutive exposures which were taken during one single night from 24 to 25th December, 2018.  The exposure time was set to 30 seconds for all data. The reason for the choice of such short and frequent observations by the proposing team was the attempted detection of stellar activity, namely p-mode oscillations. For our study here, this exceptional dataset of a very large number of exposures in a single night is ideal. First, the number of exposures is high enough to perform statistics on the RV measurements. Second, the total duration corresponds to $\simeq$ 240 m/s Doppler shift variation of telluric lines w.r.t. stellar lines, i.e. a large enough value to detect potential effects of the telluric correction on the absolute value of RV. Finally, the 4.3 hours total duration for the 200 selected exposures is too short to correspond to a detectable RV variation due to the 4.31 days period exoplanet, especially for the sequence of that particular night that corresponds to a maximum-velocity region (data points around phase 130 in Fig. 25 of \citep{Pepe2021} ) and to a RV variation of less than a few tenths of m s$^{-1}$. Because one criterion of validity of our telluric correction will be to check that the retrieved RV is constant over the period of observations, this is an important aspect.

We used the fully reduced spectra from the ESPRESSO pipeline. For each spectral order the data are given separately for each of the two parts of the image slicer and we used these distinct spectra as if they were individual orders (e.g. orders 154-155 in the next sections correspond to a unique spectral order and the two images of the slicer). For RV measurements, all available spectra are shifted to the barycentric frame. For the purpose of telluric correction we also used recorded spectra with a wavelength scale in the geocentric frame (see section \ref{telluric}).

\section{Correction of the data from telluric absorption} 
\label{telluric}
 \subsection{Principle of the correction}
  The main steps of the correction are as follows:\\

-The construction of a binary mask eliminating stellar lines to let telluric lines only  (hereafter BMT). This mask is equal to 1 in spectral intervals free of stellar lines and zero elsewhere. It can be considered as a kind of complementary mask to the BM. This mask is constructed for the orders that will be used in the next step. \\

-The determination of the atmospheric H$_{2}$O and O$_{2}$ columns present during each of the 200 exposures. These are obtained by fitting TAPAS synthetic atmospheric spectra in unmasked regions of specific appropriate orders. We checked that H$_2$O and O$_2$ are the two dominant gases in the wavelength range of ESPRESSO and that other atmospheric constituents produce negligible absorption.\\

-The computation of the product of the two H$_2$O and O$_2$ transmission spectra for each exposure and for all orders contaminated by tellurics, using the previous results for the absorbing columns. The resulting transmission spectra are convolved by the appropriate order and a wavelength dependent ESPRESSO Point Spread Function (PSF). 

-The division of the contaminated data by the computed transmission spectra.\\

 \subsection{Preparation of the binary mask eliminating stellar lines}
 
 Telluric corrections are relatively easy in the case of hot stars. Their spectra are featureless and one can fit data directly to the product of a continuum and one or several atmospheric transmission spectra to adjust the columns of absorbing gases, the telluric line Doppler shift and the PSF. The data are divided by this adjusted synthetic transmittance. For colder stars this simple procedure is often not applicable due to the huge number of stellar lines. HD40307 is relatively cool with a temperature of $\simeq$ 4980K. However, we took benefit from its low metallicity of $\simeq$ -0.3 and the high resolution of ESPRESSO, resulting in lines narrow and weak enough to allow disentangling between tellurics and stellar features in some spectral regions where the former 
 are separated from the latter, and to allow stellar continuum fitting since in those regions the stellar continuum is well recovered between the lines. This means that in such regions it is possible to mask the stellar lines and fit the spectral regions free of stellar lines to atmospheric transmission model spectra.\\ 

The masks were constructed based on a semi-automated comparison between three spectra: -a high SNR spectrum of the star obtained by stacking 10 consecutive exposures, -a TAPAS transmission of atmospheric H$_2$O and O$_2$ adapted to the Paranal site and to the date of observations, and -a stellar synthetic spectrum computed for a set of temperature, metallicity and gravity adapted to those of HD40307. This allows to select areas where telluric lines are dominant and devoid of non-negligible stellar lines. Three masks were built, respectively for orders 130-131, 146-147 (for O$_2$ lines) and orders 154-155 (for H$_2$O lines). They can be seen in Figures \ref{figure_3_h2o_removed} and \ref{figure_4_o2_removed}. The masked intervals are listed in Table \ref{Table_stellarmask}.

\subsection{Determination of \texorpdfstring{H$_2$O}{H$_2$O} or \texorpdfstring{O$_2$}{O$_2$} columns.}
 Telluric lines vary during the night in response to airmass evolution along the line-of-sight to the star and also variations of atmospheric constituents. In particular, water vapor is highly variable and may changes on timescales as short as a few minutes. Therefore it is mandatory to measure the evolution of H$_2$O and O$_2$ during the series of measurements in order to obtain a good correction.\\ 

The individual exposures were divided in sections of about 2 nm and each section was fitted in the unmasked regions to a product of a second order polynomial to represent the stellar continuum and a telluric transmittance convolved by a Gaussian PSF. The free parameters are the O$_2$ (for orders 130-131 and 146-147) or H$_2$O (for orders 154-155) columns and the coefficients of the second order polynomial. The telluric transmittance varies as a power law $T^x(H_2O)$ (resp.$T^x(O_2)$) where the factor X is proportional to the H$_2$O (resp. O$_2$) column \citep{Bertaux2014}. During the fitting procedure (which makes use of the Levenberg-Marquardt convergence scheme) the resolving power (and its associated Gaussian PSF) was fixed to a single value for the whole order estimated from Fig. 11 from \citep{Pepe2021}. Examples of fits are shown in Figures \ref{figure_3_h2o_removed} (and \ref{figure_3bis_h2o_removed_zoom}) and \ref{figure_4_o2_removed} for the first exposure. The results for the various chunks are averaged along a full order to obtain H$_2$O or O$_2$ columns temporal evolutions. \\

\subsubsection{\texorpdfstring{H$_2$O}{H$_2$O} temporal evolution}
The value of reference for the column chosen in the fitting procedure and the corresponding transmission spectrum were chosen as follows. We started with the spectrum predicted by the TAPAS website for the first exposure (25 December 2018, 00:24:00 GMT). The corresponding water vapor vertical column indicated by TAPAS COL(T) does not correspond to the column COL(P) estimated locally thanks to the infrared radiometer installed at Paranal, which monitors continuously the night sky and estimates this vertical column of H$_2$O in mm liquid equivalent (precipitable water vapor).  Such an information is included in the header of each exposure. This difference is not surprising given the strong temporal and local variability of the humidity and the fact that TAPAS interpolates temporally within ECMWF meteorological models refreshed every 6 hours and spatially within nodes that may be distant with respect to the spatial scale of the variability. We used the two values to adapt the spectrum to the Paranal conditions by elevating it to the power of the ratio COL(P)/COL(T) and attributed a coefficient X=1 to this reference spectrum.

The variation of the average fitted H$_2$O column for the 200 exposures is shown in Fig. \ref{figure_5_h2o_variation}.  It is the product of the reference column COL(P) by the fitted coefficient X. It is compared with the vertical column estimated at Paranal.  It can be seen from the comparison that the adjusted column of H$_2$O correlates well with the estimate from the infrared radiometer, in particular both have a similar amplitude of variation of about 15\% and a similar time for minimum column. They also have very similar absolute levels, showing that the chosen reference was appropriate. More precisely, the fitted value for the first exposure is X=0.93, not exactly 1 but on the same order. During some time intervals, the values deduced from the spectral adjustment differ slightly from the onsite measurements. This may probably be due to the sensitivity of radiometric data to other parameters than the molecular column.  In the subsequent correction for H$_2$O lines we made use of the spectral measurements which are the most direct. \\

\subsubsection{\texorpdfstring{O$_2$}{O2} temporal evolution}

The retrieved O$_2$ column in the line of sight (LOS) is shown in Fig. \ref{figure_6_o2_retrieved}. The chosen transmission spectrum of reference for the first exposure is exactly the one predicted by the TAPAS website and it was attributed a reference value X=1 in the fitting procedure.  The results of two series of fits are displayed, one for the order 130 which contains part of the O$_2$ $\gamma$ band near 630 nm, one for order 146 which contains part of the O$_2$ B band near 688 nm. Also shown is the evolution of the airmass for all exposures, as indicated in the headers. It can be seen that both fitted values follow very closely the temporal evolution of the airmass, which is expected. It can also be seen that there is a small difference between the two determinations. The $\gamma$ band provides a more noisy determination but at the exact expected level (X=1 for the first exposure, which happens if the fitted column is exactly the reference value for the O$_2$ column of this exposure). The B band provides a less noisy determination, however with a small relative difference of about 3 percent. The reason for this small discrepancy is unclear. It is not likely to be a cross-section error in the HITRAN database. It might be due to small differences between the modeled and actual atmospheric profiles, or, more likely, to lack of precision in the modeled PSF, in particular departures from the Gaussian approximation. During the subsequent telluric line removal, and in view of these results, we decided to use the B band result for the absolute value of X, because the B band lines are stronger, and for the temporal evolution we have chosen to use quite simply the airmass as the multiplicative factor for the O$_2$ vertical column .\\

\subsection{Correction for telluric lines based on the \texorpdfstring{H$_2$O}{H$_2$O} and \texorpdfstring{O$_2$}{O2} adjusted columns.}

\begin{figure*}
\centering
\includegraphics[width=\hsize]{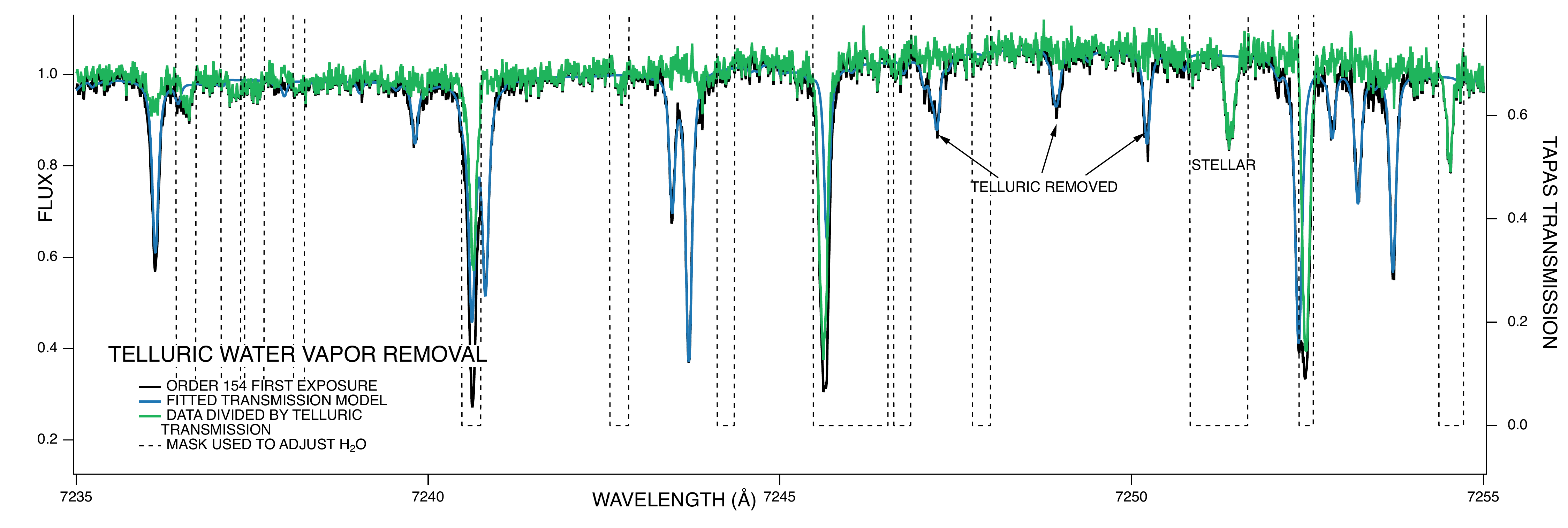}
\caption{Illustration of the procedure used to determine the H$_2$O column for each exposure (here exposure 0 and order 154) and the telluric H$_2$O removal. The mask of stellar lines BMT is shown in black dashed lines. The data and adjusted TAPAS transmittance are shown in black and light blue respectively. The division of the data by the fit is shown in green. The correction works for isolated telluric lines and also for blended telluric and stellar lines (e.g. at 7240 A, see figure \ref{figure_3bis_h2o_removed_zoom}.}
\label{figure_3_h2o_removed}
\end{figure*}

\begin{figure*}
\centering
\includegraphics[width=\hsize]{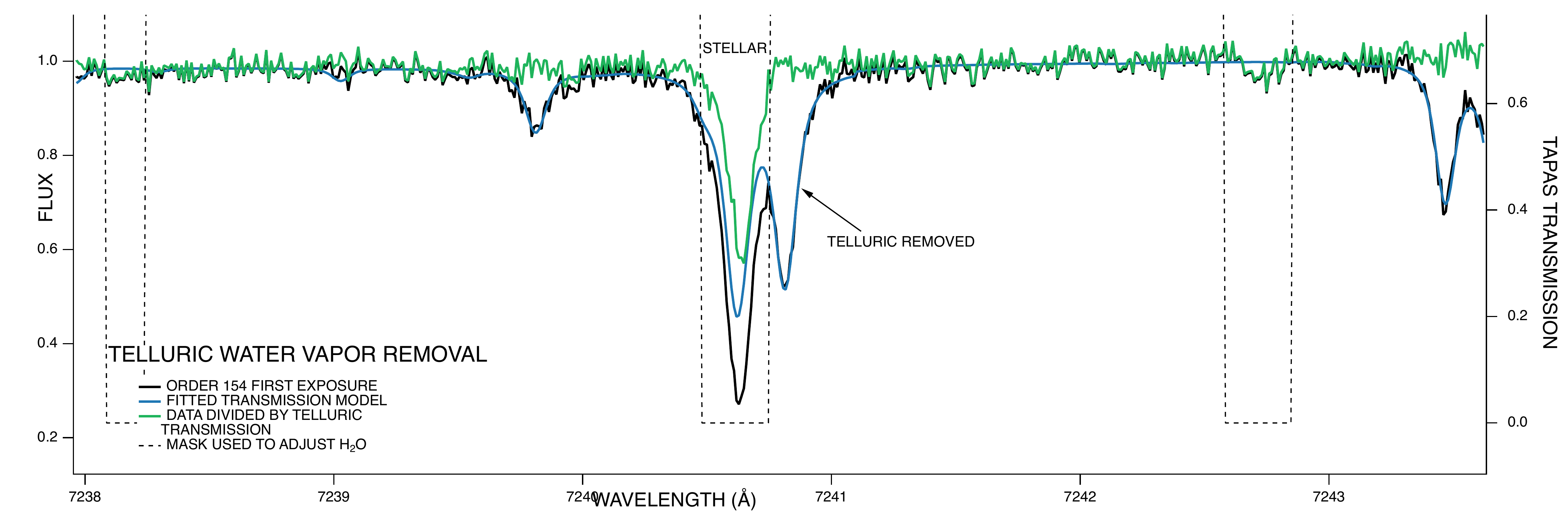}
\caption{Zoom on Figure \ref{figure_3_h2o_removed}. Even stellar lines heavily contaminated by H$_2$O lines may be kept for RV retrieval after correction from an adjusted telluric transmission. The mask of stellar lines BMT is shown in black dashed lines.}
\label{figure_3bis_h2o_removed_zoom}
\end{figure*}

\begin{figure*}
\centering
\includegraphics[width=\hsize]{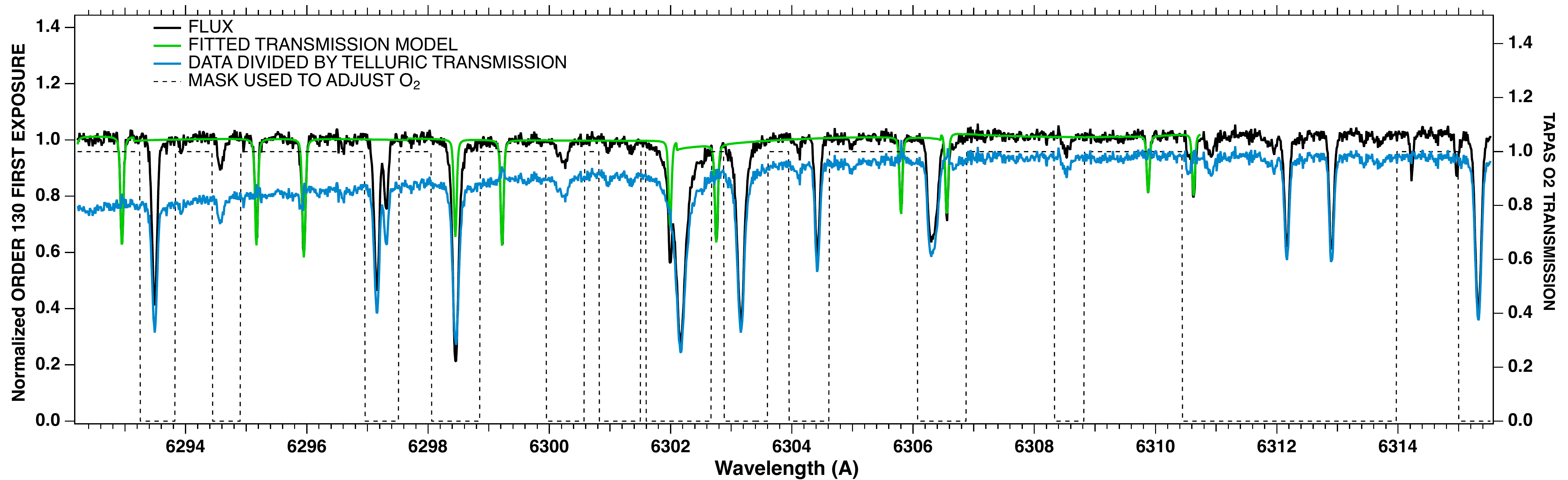}
\caption{Illustration of the procedure used to determine the O$_2$ column for each exposure (here exposure 0 and order 130). The mask of stellar lines BMT is shown in black dashed line. The normalized data and the fitted TAPAS transmission are shown in black and green respectively. The division of the data by the fit is shown in blue.}
\label{figure_4_o2_removed}
\end{figure*}

\begin{figure}
\centering
\includegraphics[width=\hsize]{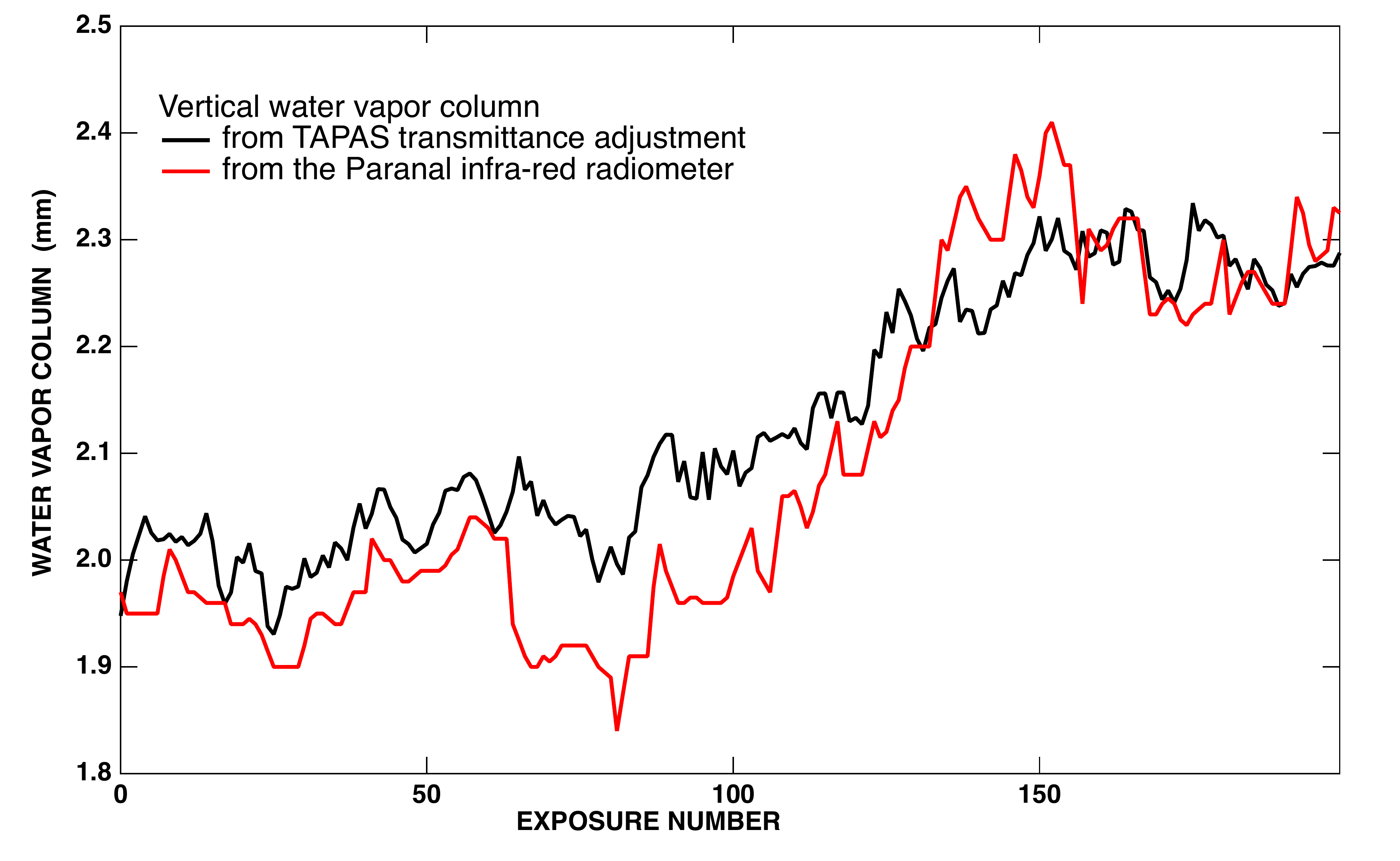}
\caption{Comparison between the vertical column of water vapor adjusted through spectral fitting (black curve) and the one estimated from the infra-red monitor at Paranal (in red) for the 200 consecutive exposures.}
\label{figure_5_h2o_variation}
\end{figure}

\begin{figure}
\centering
\includegraphics[width=\hsize]{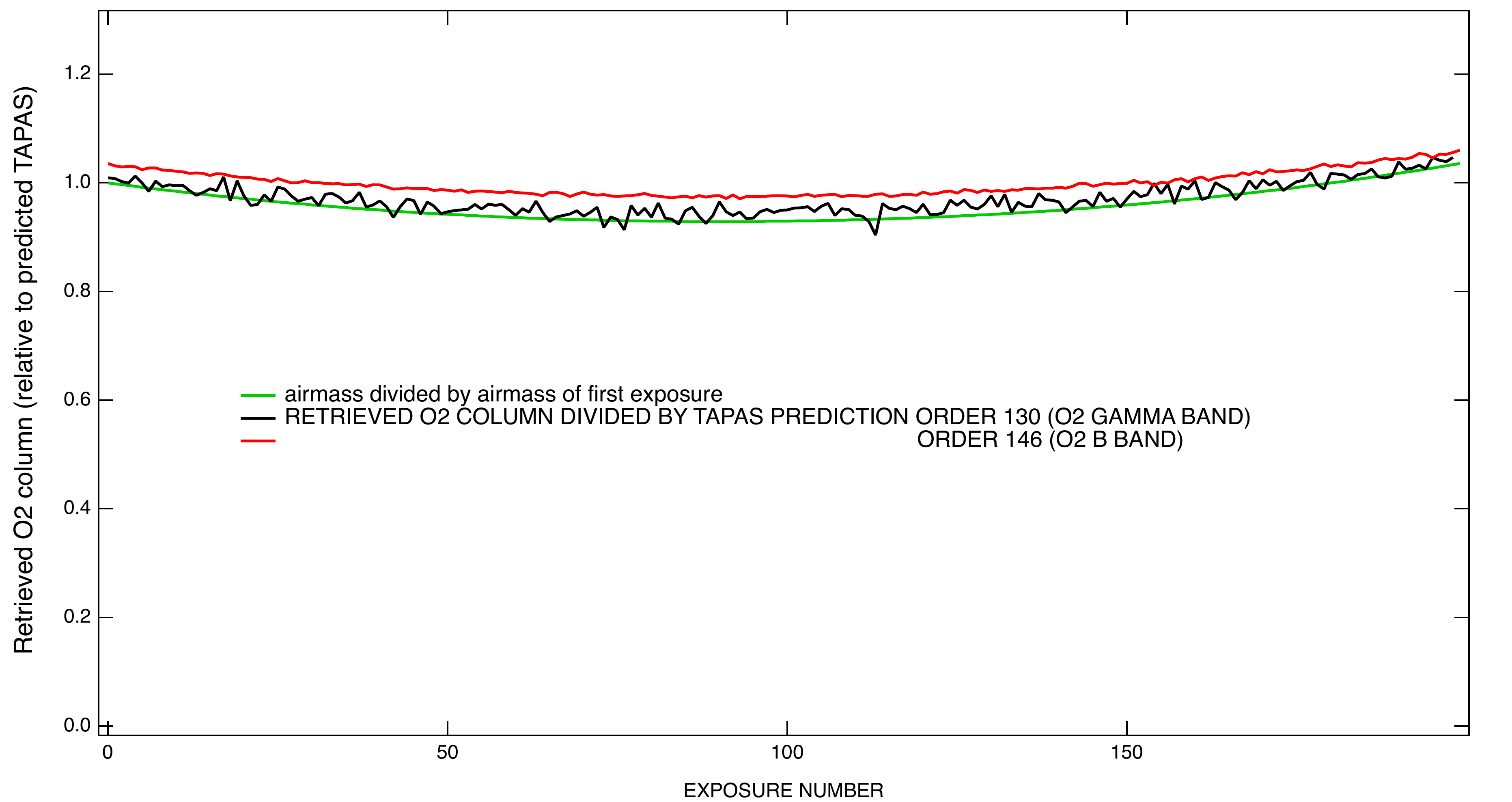}
\caption{Comparison between the column of O$_2$ adjusted through spectral fitting of the B band and the $\gamma$ band (black and red curves) and the air mass (green curve). The retrieved columns of O$_2$ follow quite well the air mass, which means that, unlike the H$_2$O variations, there is no significant variation of the O$_2$ vertical column.}
\label{figure_6_o2_retrieved}
\end{figure}

\begin{figure}
\centering
\includegraphics[width=\hsize]{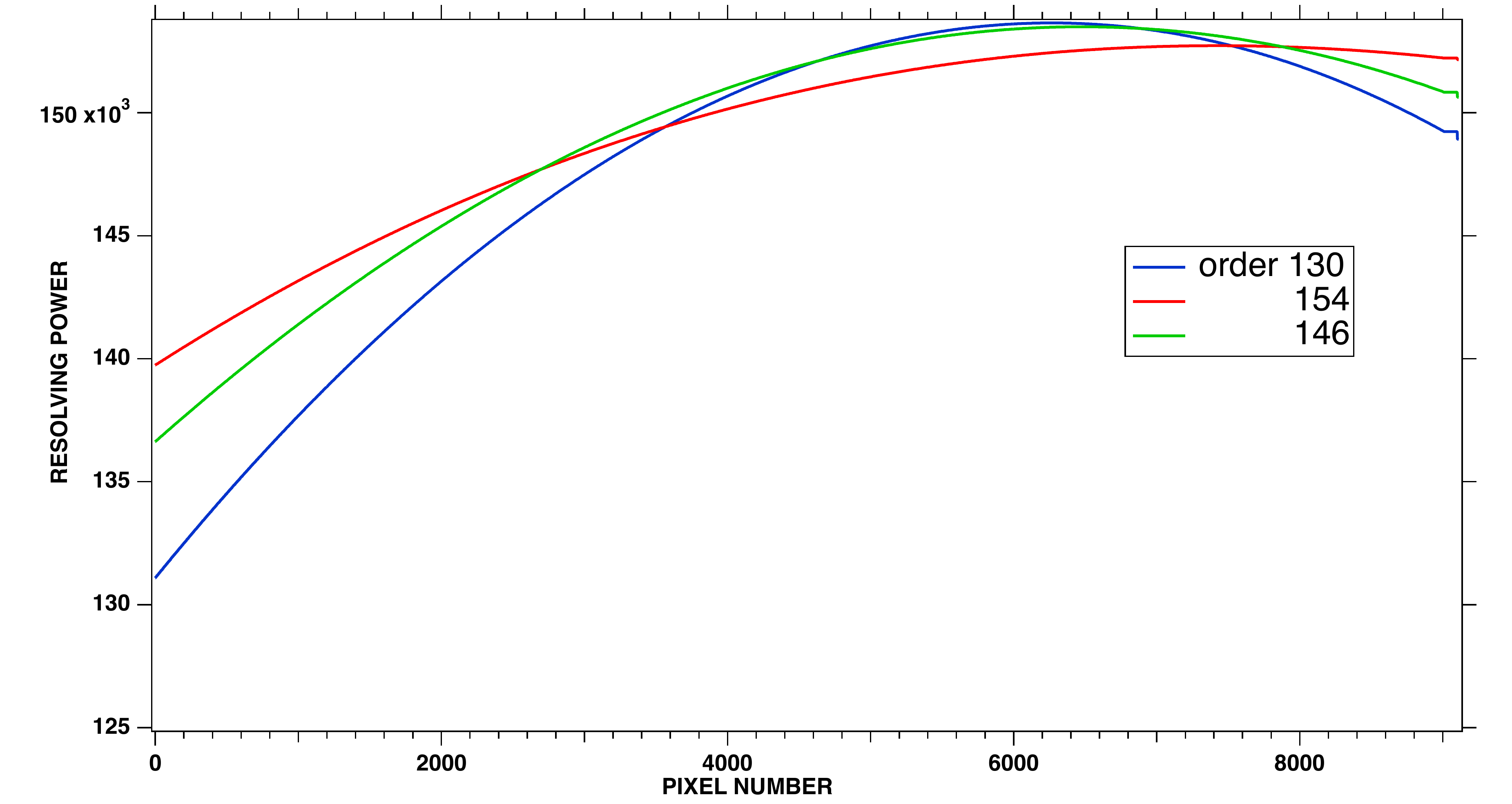}
\caption{Spectral resolution as a function of pixel number, based on the digitization of Figure 11 of \citep{Pepe2021}. Shown are three examples, for three orders of the red arm.}
\label{figure_7_resolvingpower}
\end{figure}

 The next and final step of the correction procedure is, for the full wavelength range, the division of the data by a synthetic transmission computed for each exposure for the adjusted H$_2$O and O$_2$ values and adapted to the instrument resolution in an optimal way. The spectral resolution is not constant along an order and from order to the other, as it is well known. Fortunately the information on the resolution is available from Figure 11 of \cite{Pepe2021}. We have digitized this figure and the associated color scale and derived the spectral resolution as a function of order and wavelength. The resolution for the three orders used for the O$_2$ and H$_2$O estimate are displayed in Figure \ref{figure_7_resolvingpower}, showing amplitudes of variations on the order of 15 percent along an order. We used the results of this digitization to convolve the atmospheric transmission by the correct wavelength-dependent resolution, i.e. taking into account the true PSF for each order and each wavelength. The last operation, namely the division of the data by the convolved synthetic transmission is not fully mathematically correct. In principle the recorded spectrum should be de-convolved first, then divided by the transmission, and the result of the division should be re-convolved \citep{Bertaux2014}. However, for a high resolving power the first simplified approach is a good approximation. Examples of the divisions are shown in Figures \ref{figure_3_h2o_removed}, \ref{figure_3bis_h2o_removed_zoom} and \ref{figure_4_o2_removed} for H$_2$O and O$_2$. 

For the whole correction we assumed that the PSF is exactly Gaussian and we only varied its width according to the wavelength and the order. We did it for the following reasons. First, in the case of hot stars with smooth continua, it is possible to deconvolve isolated telluric lines to infer the PSF shape, however, here in the case of this K star the result of the deconvolution would be somewhat hazardous. Second, there are no indications of departures from a Gaussian shape in the ESPRESSO documents and, to our knowledge, no mentions of departures in articles based on ESPRESSO data. Third, and most important, significant departures would introduce features at the locations of the wings of the corrected lines that would repeat along all lines, especially strong for a deep absorption, and we did not notice any specific, repetitive shape in the residuals in the wings that would be such an indicator. On the contrary, the residuals, when present, do vary from one line to another.

 \begin{table*}

\caption{Stellar mask for orders 130/131, 146/147 and 154/155. Wavelengths are in stellar rest frame and in vacuum. For each order the two limits of the masked region are listed. }             % title of Table
\label{Table_stellarmask}      % is used to refer this table in the text
\centering                          % used for centering table
\begin{tabular}{|c c|| c c|| c c| }        % centered columns (4 columns)
\hline\hline                 % inserts double horizontal lines
\multicolumn{2}{c}{Order 130}  &  \multicolumn{2}{c}{Order 146} & \multicolumn{2}{c}{Order 154}\\    % table heading  \\
\hline                        % inserts single horizontal line
Beginning & End &Beginning & End &Beginning & End \\
\hline
6266.54	&	6267.09	&	6829.64	&	6830.88	&	7150.50	&	7150.87	\\
6267.54	&	6268.23	&	6831.31	&	6832.02	&	7151.57	&	7152.37	\\
6270.10	&	6270.83	&	6833.66	&	6834.25	&	7152.76	&	7153.13	\\
6271.45	&	6272.28	&	6834.71	&	6835.21	&	7153.25	&	7153.88	\\
6272.64	&	6275.40	&	6838.34	&	6839.06	&	7153.90	&	7154.26	\\
6275.65	&	6277.28	&	6840.10	&	6848.95	&	7155.06	&	7155.77	\\
6278.95	&	6279.27	&	6850.08	&	6867.81	&	7156.12	&	7157.10	\\
6281.28	&	6281.70	&	6872.62	&	6872.92	&	7157.16	&	7158.03	\\
6282.00	&	6282.76	&	6873.10	&	6874.46	&	7158.45	&	7158.84	\\
6283.31	&	6283.60	&	6875.97	&	6876.31	&	7159.57	&	7159.95	\\
6284.03	&	6284.67	&	6877.07	&	6877.58	&	7160.24	&	7160.63	\\
6286.67	&	6287.08	&	6878.03	&	6878.43	&	7161.93	&	7162.48	\\
6287.55	&	6288.04	&	6881.27	&	6881.57	&	7162.58	&	7162.96	\\
6289.86	&	6290.23	&	6882.01	&	6885.28	&	7163.37	&	7163.76	\\
6292.05	&	6293.01	&	6887.38	&	6888.42	&	7164.10	&	7164.48	\\
6294.31	&	6294.89	&	6896.07	&	6896.51	&	7164.95	&	7165.59	\\
6295.50	&	6295.97	&	6899.70	&	6900.46	&	7165.97	&	7166.83	\\
6298.02	&	6298.58	&	6902.67	&	6903.36	&	7166.87	&	7167.25	\\
6299.12	&	6299.92	&	6904.46	&	6904.99	&	7167.32	&	7167.69	\\
6301.01	&	6301.64	&	6907.02	&	6909.09	&	7168.73	&	7169.11	\\
6301.88	&	6302.57	&	6909.99	&	6910.26	&	7170.87	&	7171.46	\\
6302.66	&	6303.73	&	6910.87	&	6914.11	&	7171.88	&	7172.26	\\
6303.94	&	6304.67	&	6914.92	&	6915.62	&	7172.27	&	7173.02	\\
6305.01	&	6305.68	&	6915.88	&	6916.78	&	7175.50	&	7176.50	\\
6307.13	&	6307.94	&	6917.59	&	6917.88	&	7176.59	&	7176.96	\\
6309.39	&	6309.89	&	6918.16	&	6919.04	&	7177.70	&	7178.27	\\
6311.50	&	6315.04	&	6919.69	&	6920.00	&	7178.70	&	7179.11	\\
6316.06	&	6324.64	&	6920.33	&	6920.59	&	7180.54	&	7180.91	\\
6326.72	&	6329.78	&	6921.04	&	6921.77	&	7181.54	&	7182.41	\\
6330.06	&	6330.45	&	6921.99	&	6922.19	&	7182.98	&	7183.34	\\
6331.54	&	6335.06	&	6923.05	&	6923.15	&	7183.74	&	7184.10	\\
6336.50	&	6341.04	&	6924.15	&	6924.29	&	7185.59	&	7185.93	\\
6345.60	&	6346.22	&	6925.49	&	6930.24	&	7186.69	&	7187.04	\\
6351.05	&	6351.42	&		&		&	7188.80	&	7189.81	\\

\hline                                   %inserts single line
\end{tabular}
\end{table*}

\section{Standard and extended binary masks to be used in the classical Cross-Correlation Function method}
\label{RV}
In order to quantify the quality gain obtained by correcting the spectra from telluric absorptions, we need to compare the retrieved RVs with exactly the same algorithm, applied either to the non-corrected data (basically the wavelengths regions non contaminated by tellurics), and to the corrected data (including in addition  wavelength regions corrected from tellurics). 
Actually, there are several types of RV retrieval algorithms, but for our purpose, we do not need to use the ultimate best algorithm, as long as we use exactly the same one for the two cases.

The historical and classical approach of RV retrieval is the Cross-Correlation Function (CCF) with a binary mask \citep{Baranne1996,Pepe2002}, which is actually used in the official ESPRESSO pipeline for the official RV retrieval. It has the advantage of being robust. One of the recent papers describing in details this method is \cite{Lafarga2020}, from which we borrowed some details when coding our own version of this algorithm.

For processing the non-corrected data, we used the standard binary mask of ESPRESSO corresponding to the spectral type K2.5  of HD40307. This mask is made of a series of wavelengths in air and the relative depth of the star line (called the contrast), which is represented in Fig.\ref{figure_1_mask} as a function of wavelength. When compared to the model transmission computed by TAPAS (Fig. \ref{figure_1_mask}), it is clear that this standard binary mask is avoiding regions seriously contaminated by atmospheric H$_2$O and O$_2$ lines. 
We selected all the lines of the standard mask with a contrast >0.2, which constituted the BM1 mask that was used to process the non-corrected data. A second mask, BM2, was built by adding to the BM1 mask a series of new lines, in regions contaminated by telluric absorptions, constituting a mask BMc. In fact, there are many orders where there were no lines in  BM1: 118, 136, 148, 150, 152, 154, 156, and 164 (and their odd numbers twins). 
In order to find the new lines of BM2=BM1+BMc, we stacked together the 200 spectra after telluric correction, and to this stacked spectrum we assigned the wavelengths of the exposure 100, in the reference frame of ESPRESSO. Then, for a set of observed stellar spectral features, we determined the precise wavelength  (by a Gaussian fit) of the minimum and kept those lines which had a contrast >0.2. In Figure \ref{figure_2_number_lines} are plotted the number of lines per order for the two masks BM1 and BM2 as a function of order number. For some orders (116 to 123, 128 to 148, 158 to 159) the number of lines from BM1 was not zero but small; the lines were kept for BMc but their wavelengths were taken also from the stacked spectrum (after telluric correction). In Table \ref{Tab_numberoflines} are listed the numbers of initial (BM1), additional selected (BMc) and total lines of the binary mask. There is only one more line for the blue arm, and 724 more lines for the red arm, added to the 1060 lines of the original mask BM1. Actually, for the analysis of the time series of measurements, we dropped the last 6 orders (A band of O$_{2}$ and weak signal); the selected orders for the red arm contain 696 lines more than the 1050 lines of BM1. The gain in RV measurement precision is entirely due to the use of those 696 extra-lines of BMc mask. Table \ref{Table_maskorder130} lists the BMc lines for order 130. The full list of additional lines will be available on request.

\begin{table}
\caption{Number of lines in the binary mask}             % title of Table
\label{Tab_numberoflines}      % is used to refer this table in the text
\centering                          % used for centering table
\begin{tabular}{c c c c}        % centered columns (4 columns)
\hline\hline                 % inserts double horizontal lines
Binary Mask & BM1  & BMc & BM2=BM1+BMc \\    % table heading  \\
\hline                        % inserts single horizontal line
All orders 40-169	& 3654 & 	725	&4379\\
Blue arm 40-89&	2594&	1&	2595\\
Red arm 90-169&	1060&	724	&1784\\
Red arm 90-163&	1050&	696	&1746\\
\hline                                   %inserts single line
\end{tabular}
\end{table}

\begin{table}
\caption{Binary mask lines for order 130/131. Wavelengths are in vacuum.}             % title of Table
\label{Table_maskorder130}      % is used to refer this table in the text
\centering                          % used for centering table
\begin{tabular}{c c c }        % centered columns (4 columns)
\hline\hline                 % inserts double horizontal lines
 Wavelength in vacuum & Contrast  & FWHM ms$^{-1}$  \\    % table heading  \\
\hline                        % inserts single horizontal line
 6266.8465	&	0.8	&	4486.20\\
6268.0325	&	0.3	&	2602.77\\
6270.5443	&	0.8	&	12388.95\\
6271.9383	&	0.5	&	2677.01\\
6272.9907	&	0.3	&	2519.84\\
6276.3673	&	0.5	&	3034.65\\
6281.4607	&	0.3	&	3953.09\\
6282.3338	&	0.8	&	3967.68\\
6284.3536	&	0.6	&	12500.92\\
6286.8781	&	0.6	&	2914.92\\
6292.6845	&	0.4	&	2634.84\\
6294.5440	&	0.6	&	3009.47\\
6298.2113	&	0.5	&	2847.12\\
6298.3656	&	0.2	&	1939.91\\
6299.5157	&	0.7	&	3613.13\\
6303.2231	&	0.6	&	3603.95\\
6304.2168	&	0.6	&	3534.69\\
6305.4806	&	0.4	&	2433.84\\
6307.3757	&	0.8	&	17150.33\\
6313.2248	&	0.4	&	2445.00\\
6313.9634	&	0.4	&	2383.63\\
6316.3841	&	0.7	&	4334.12\\
6317.0346	&	0.5	&	2684.57\\
6317.5363	&	0.4	&	2565.93\\
6319.7486	&	0.9	&	10887.55\\
6320.4338	&	0.2	&	2874.01\\
6320.9591	&	0.2	&	3435.50\\
6324.4158	&	0.8	&	4638.91\\
6329.3274	&	0.4	&	2685.68\\
6331.8214	&	0.6	&	2797.14\\
6332.5758	&	0.2	&	2317.80\\
6337.0623	&	0.8	&	5807.31\\
6337.8318	&	0.3	&	2552.29\\
6338.5565	&	0.7	&	4091.04\\
6340.6073	&	0.3	&	2541.33\\
6340.8357	&	0.3	&	2592.87\\
6345.8836	&	0.6	&	3119.11\\
6356.7665	&	0.8	&	5692.01\\    

\hline                                   %inserts single line
\end{tabular}
\end{table}

\begin{figure*}
\centering
\includegraphics[width=0.95\hsize]{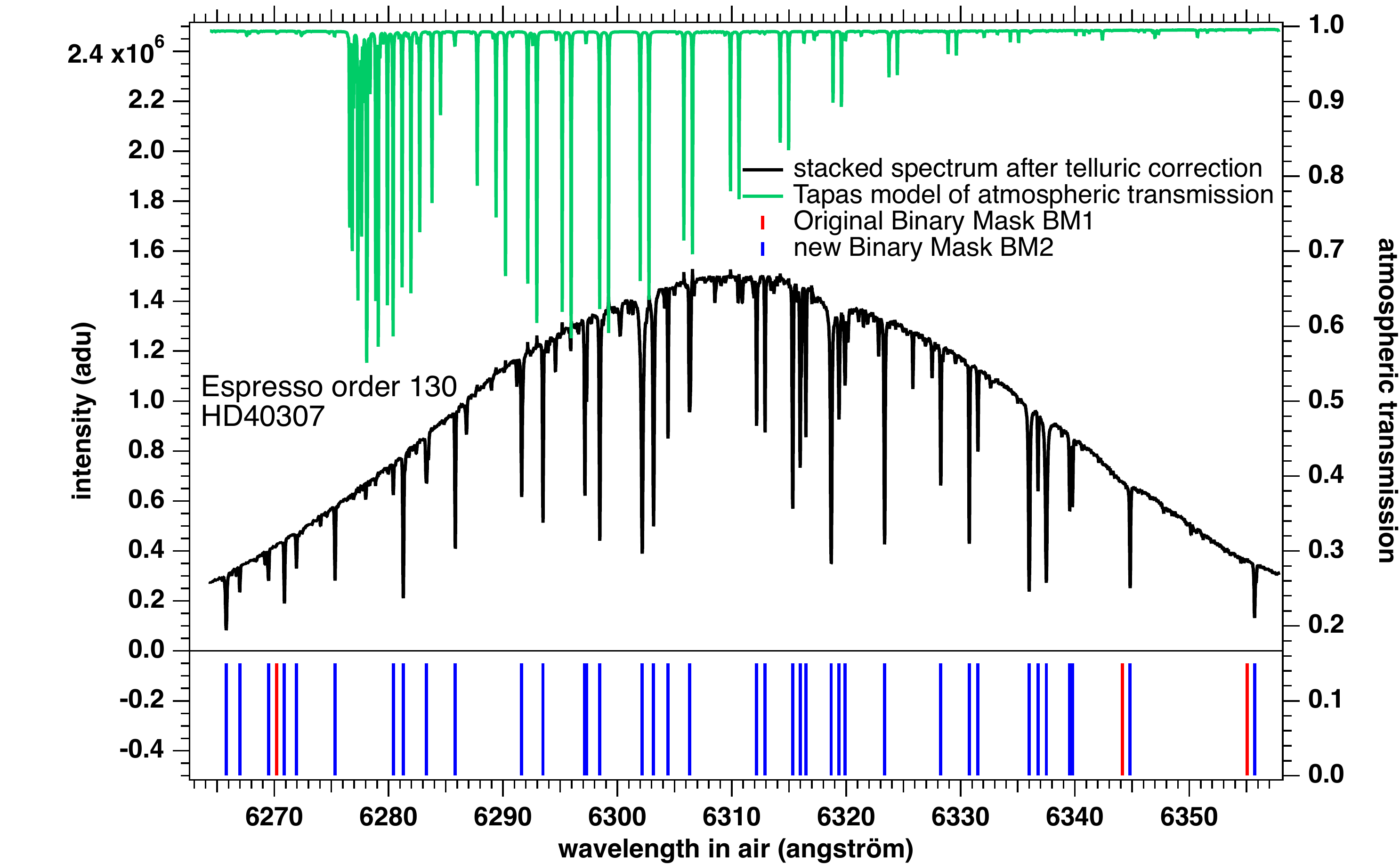}
\caption{Augmented mask of order 130. The intensity of the stacked spectrum (high SNR), after correction of telluric absorptions, is shown in black (left scale). The atmospheric transmission computed from TAPAS (green, right scale) displays the $\gamma$ band of O$_2$. The original Binary Mask BM1 contains only 3 lines, which wavelengths are indicated at bottom by red vertical bars. They are avoiding the O$_2$ contaminated region. A new Binary Mask BM2 containing 36 lines is made from the observed stacked spectrum with stellar lines which depths are $\geq$ 0.2. The wavelength assigned to the stacked spectrum is the one of exposure Number 100, in the middle of the series of exposures. The 3 red lines of BM1 are seen shifted to the left from their blue line counterpart detected in the stacked spectrum.}
\label{figure_8_order_130_mask}
\end{figure*}

On figure \ref{figure_8_order_130_mask} is represented the order 130 of the stacked spectrum, as well as the TAPAS model transmission and the position of the lines represented by vertical bars of different colors for BM1 and BMc. This order is contaminated by the $\gamma$ band of di-oxygen O$_2$, and as a result there are only three lines in this order in the BM1 mask (position indicated by vertical red lines at bottom). We could find 33  more lines for the BMc lines (vertical blue lines) giving a total of 36 lines for this order.

\begin{figure}
\centering
\includegraphics[width=0.9\hsize]{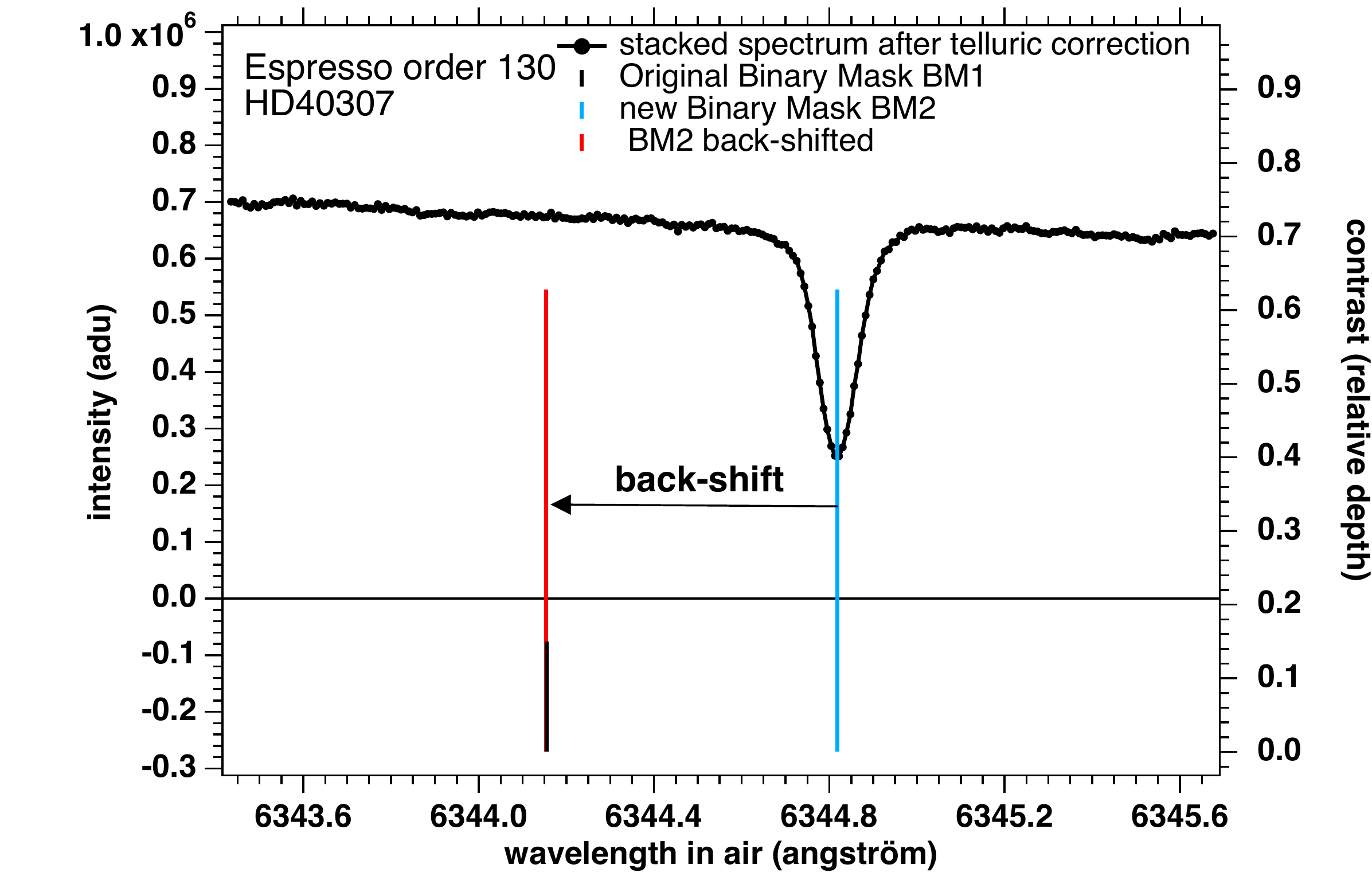}
\caption{Adjustment of the two masks. The vertical blue line gives the wavelength position of the center of one stellar line of the observed spectrum, after telluric correction and stacking of all exposures. Its height is the contrast of the line (relative depth, 0.64). The red line represent the same line, back-shifted to account for the radial velocity Vrad$_{0}$ of the star. The velocity Vrad$_{0}$ was determined with the original mask BM1 and data from orders 92 to 115. As a result, the position of the red line coincides well (but not perfectly) with the corresponding line of the original BM1 mask (vertical bar, bottom).}
\label{figure_9_order_130_mask_line}
\end{figure}

\begin{figure*}
\centering
\includegraphics[width=0.9\hsize]{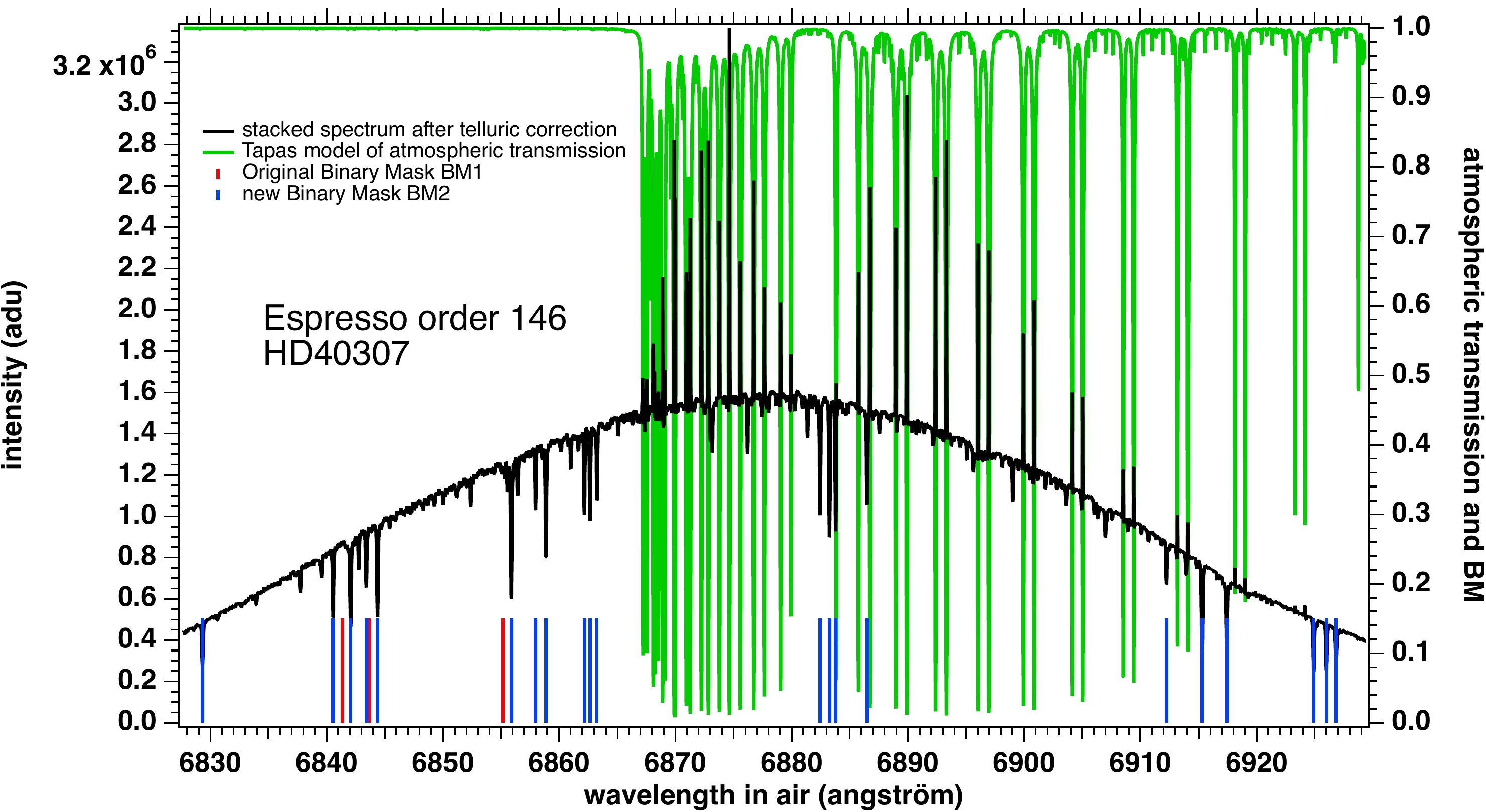}
\caption{Same as Figure \ref{figure_8_order_130_mask} for the order 146, heavily contaminated by deep, narrow lines of the O$_{2}$ B band. The corrected spectrum exhibits conspicuous residuals (see text) at the location of the strongest absorption, however we could add 18 lines in the new mask in addition to the initial three lines.}
\label{figure_10_order_146_mask}
\end{figure*}

Figure \ref{figure_9_order_130_mask_line} is a zoom of figure \ref{figure_8_order_130_mask} around a particular line at 6344.8 \AA~ which was already in BM1, which position is indicated by the black vertical line at bottom.  The blue line is the contrast (right scale) of the line, centered on the observed spectral line, displaced by Doppler shift from the BM1 line. But to serve as a line in a binary mask we need to shift this new line back to a reference system with RV=0. Therefore, to operate this back-shift we have to select precisely the RV of the star HD40307. We choose Vrad$_{0}$= 31381.572 ms$^{-1}$, as determined with the mask BM1 with orders 92 to 115. When this line is back-shifted, it comes to the position of the red vertical bar, very near the original BM1 line. All new lines from telluric corrected stacked spectrum have been back-shifted by the same stretch factor (the Doppler shift is actually a stretch) to constitute the mask BMc, added to BM1 to form BM2. 

In Figure \ref{figure_10_order_146_mask} we show the extreme case of the order 146 which is heavily contaminated by the O$_{2}$ B band. The correction is poor at the location of the strongest absorption lines, and strong residual spikes are visible. A better correction would require a detailed adjustment of the exact shape of the PSF, beyond the scope of this work. This order is also characterized by a low number of stellar lines. Still, we could add 18 lines to the binary mask. To do so, we avoided the regions with poor correction.  

In the CCF scheme described by \cite{Lafarga2020} that we adopted, each order is treated separately. The CCFs of all the lines of one order are added together with a weight attached to each line, then the total CCF is fitted by a Gaussian which position of center (the minimum) is the radial velocity for this order. The uncertainty $\sigma_{ord}$ attached to this determination is the one returned by the Gaussian fit routine. 
Once the RVs for each order are determined, they are averaged together, accounting in a standard way their uncertainties: weighted average with weights =1/$\sigma^{2}_{ord}$, and the error for the exposure $\sigma_{exp}$ is such that 1/$\sigma_{exp}^{2}$=$\Sigma$ 1/$\sigma^{2}_{ord}$.

While \cite{Lafarga2020} took the ratio contrast/width as a weight attached to each line of a BM, we took only the contrast of each line as its weight for its CCF. The reason is that the mask BM1 contains only the contrast and not the width. Since we wanted to use exactly the same algorithm for comparing the RV results with BM1 and BM2, we were forced to use only the contrast for the two cases. We have conducted a limited exercise of comparison with the two versions of the weight, either the contrast only, or the contrast/width. We found out a very modest decrease (2.9 \%) of the uncertainty and $\simeq$ 3\% for the spread of all RV measurements (standard deviation) among the series of 200 exposures. Therefore, while clearly it would be slightly better to use the weight= contrast/width, using only the contrast will not be detrimental for the comparison between the use of BM1 applied to original spectra and BM2 applied to telluric corrected spectra. 

The CCF of one line of the BM with one observed line is done by computing the signal contained in a boxcar centered on the BM line, and displaced by a variable RV (by steps of 492 ms$^{-1}$, the mean size of one spectel) around the observed minimum. Following the recommendation of \cite{Lafarga2020}, we chose a size of the boxcar width about equal to one spectel of the observed spectrum. As a result, the CCF is very similar to the observed line, because both the boxcar width and the step of the RV grid are of the size of one spectel. It has also the advantage of providing independent points of the CCF. Indeed, the Gaussian fit algorithm used in Igor language (but this is the same for most coding languages having coded the Numerical Recipe library \citep{Press92} is assuming that all points to be fitted are independent, and the error bar returned by the fit routine rely on this hypothesis. This would not be the case, either if the boxcar width would be larger than one pixel, or if the RV grid step was smaller than one spectel. We found out that one important parameter of the Gaussian fit to the total CCF for one order is the number of CCF steps over which is fitted the CCF, because it was found that the exact position of the CCF minimum depends substantially on this number. It was fixed to 10 points in all the presented results. 

The blaze effect is quite conspicuous on figure \ref{figure_8_order_130_mask}. We did not attempt to correct the blaze before performing the CCF. When fitting a tilted (from constant) signal, it introduces a bias on the position of the minimum. However, we determined that with the observed slopes due to the ESPRESSO blaze, the bias does not exceed about 12 ms$^{-1}$. This systematic bias is constant along all exposures for one particular line and does not increase the spread of all measurements, nor the (shot noise associated) uncertainty on RV. Furthermore, being of opposite sign on the two sides of the order, they will more or less compensate when doing the CCF for a full order. Also, we found out that for some lines of the BM1 mask (probably those determined from laboratory measurements of transitions), the bias was much larger, of the order of $\simeq$ 100 ms$^{-1}$, due to convection and granulation, as pointed out by \cite{Gonzales2020}, therefore much larger than the biasing blaze effect. 

Because in this CCF algorithm there are several adjustable parameters (as described above), we do not expect to find, for each exposure, exactly the same RV value as the one contained in the archive product of HD40307 (for which the exact values of the parameters are different from ours and unknown to us). But this is not our objective. Our purpose is to compare the RV series results obtained with BM1 and BM2, with exactly the same algorithm and its parameters. Also, we have not used all the lines of the original mask, keeping only those with a contrast larger than 0.2.

As suggested by our reviewer, we also computed another estimator of the error on RV, based on quality factor and photon-noise as first devised by \cite{Connes1985} and revived by \cite{Bouchy2001}. To do so, we used the approach from \cite{Boisse2010}, in which authors consider a CCF for one order as an averaged noise-free spectral line. It provides an estimate of the minimal error. We detail this approach and present our calculations in Appendix \ref{appendix}.

\begin{figure*}
\centering
\includegraphics[width=0.9\hsize,height=0.35\hsize]{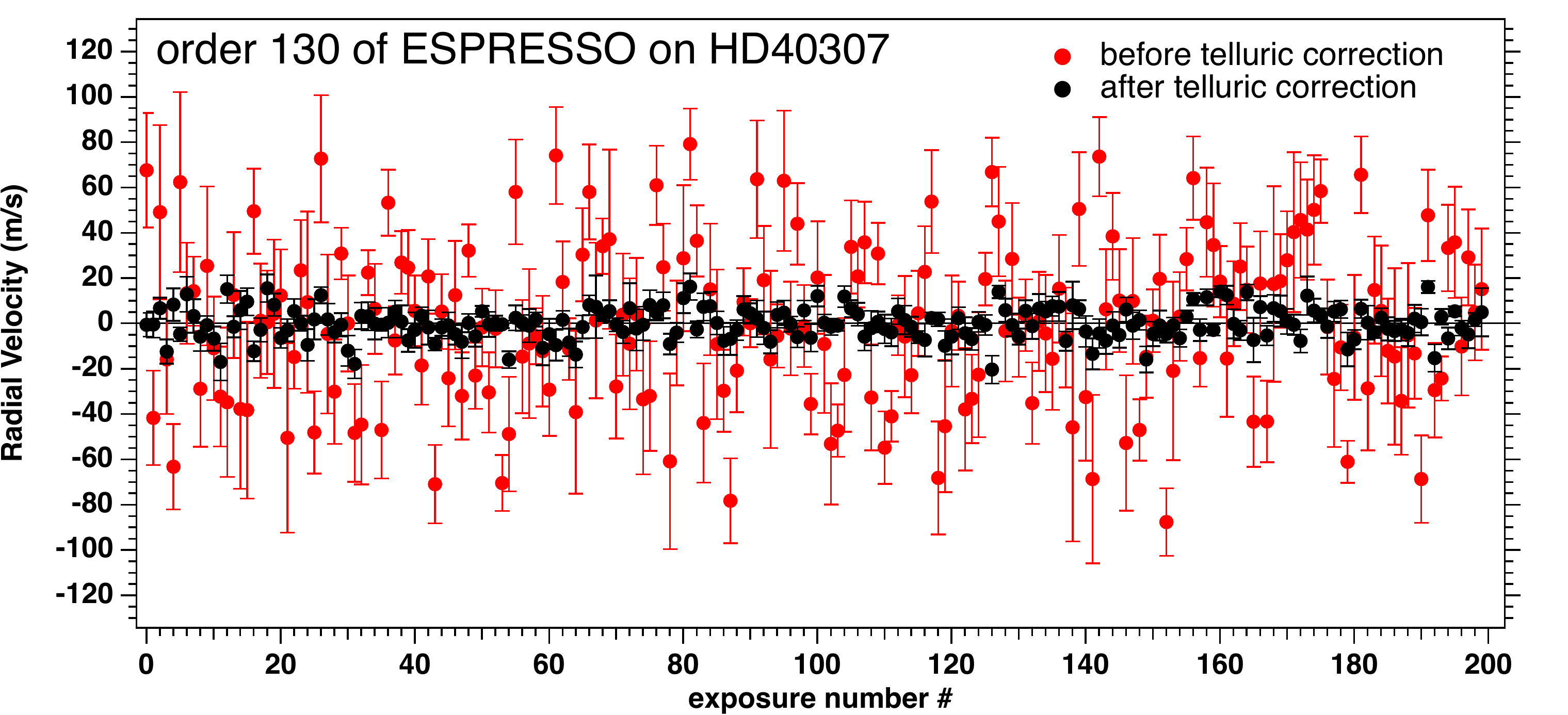}
\caption{Radial velocity of star HD40307 for order 130, which is contaminated by O$_{2}$. In red - result with the standard binary mask BM1 before telluric correction; in black – result with our new binary mask BM2  after telluric correction. The average radial velocities of 31372.4 and 31384.2 ms$^{-1}$ were subtracted respectively from the series before and after correction.}
\label{figure_11_comparisonorder130}
\end{figure*}

\begin{figure*}
\centering
\includegraphics[width=0.9\hsize,height=0.35\hsize]{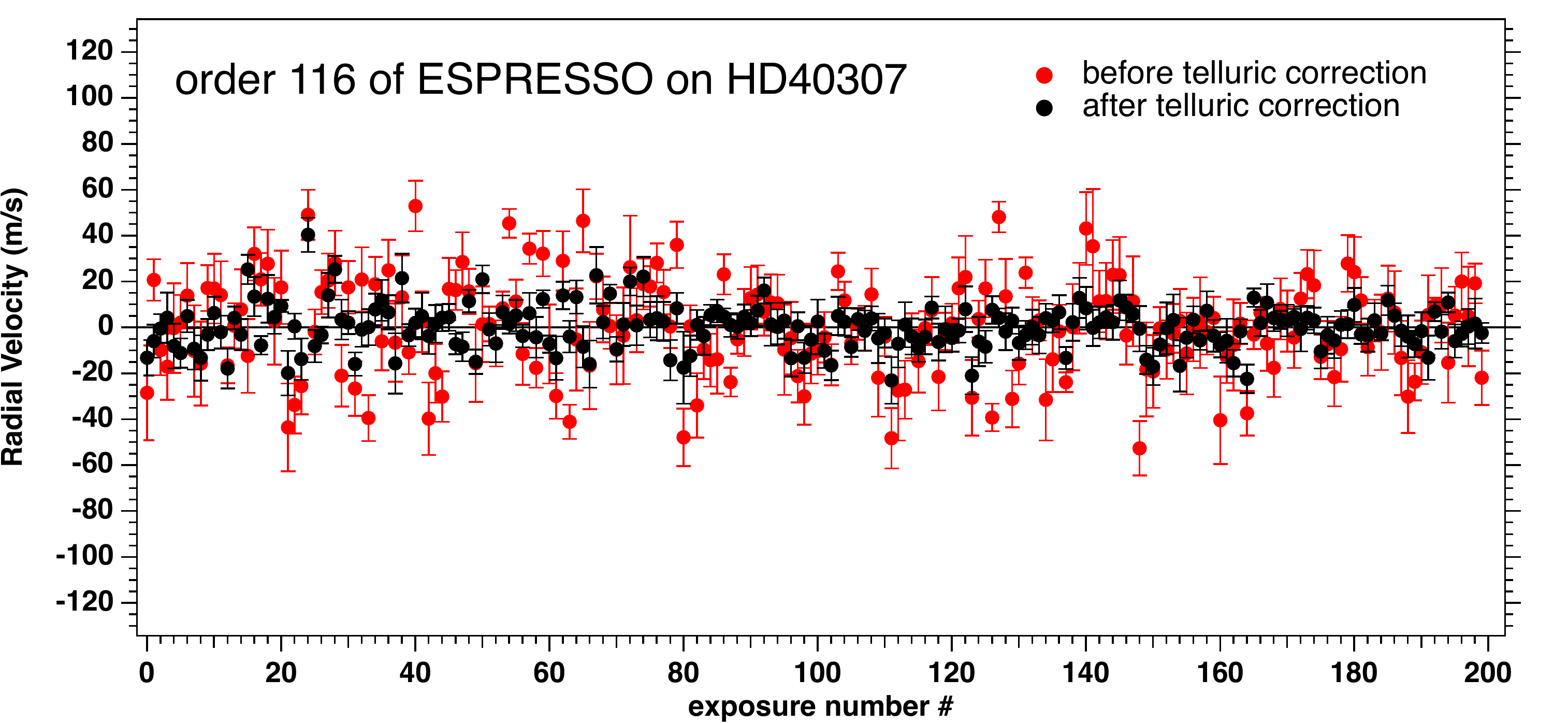}
\caption{Radial velocity of star HD40307 for order 116, which is contaminated by H$_2$O. In red - result with the standard binary mask BM1  before telluric correction; in black – result with our new binary mask BM2  after telluric correction. The average radial velocities of 31386.4 and 31388 ms$^{-1}$ were subtracted respectively from the series before and after correction.}
\label{figure_12_comparisonorder116}
\end{figure*} 

\begin{figure*}
\centering
\includegraphics[width=0.9\hsize,height=0.35\hsize]{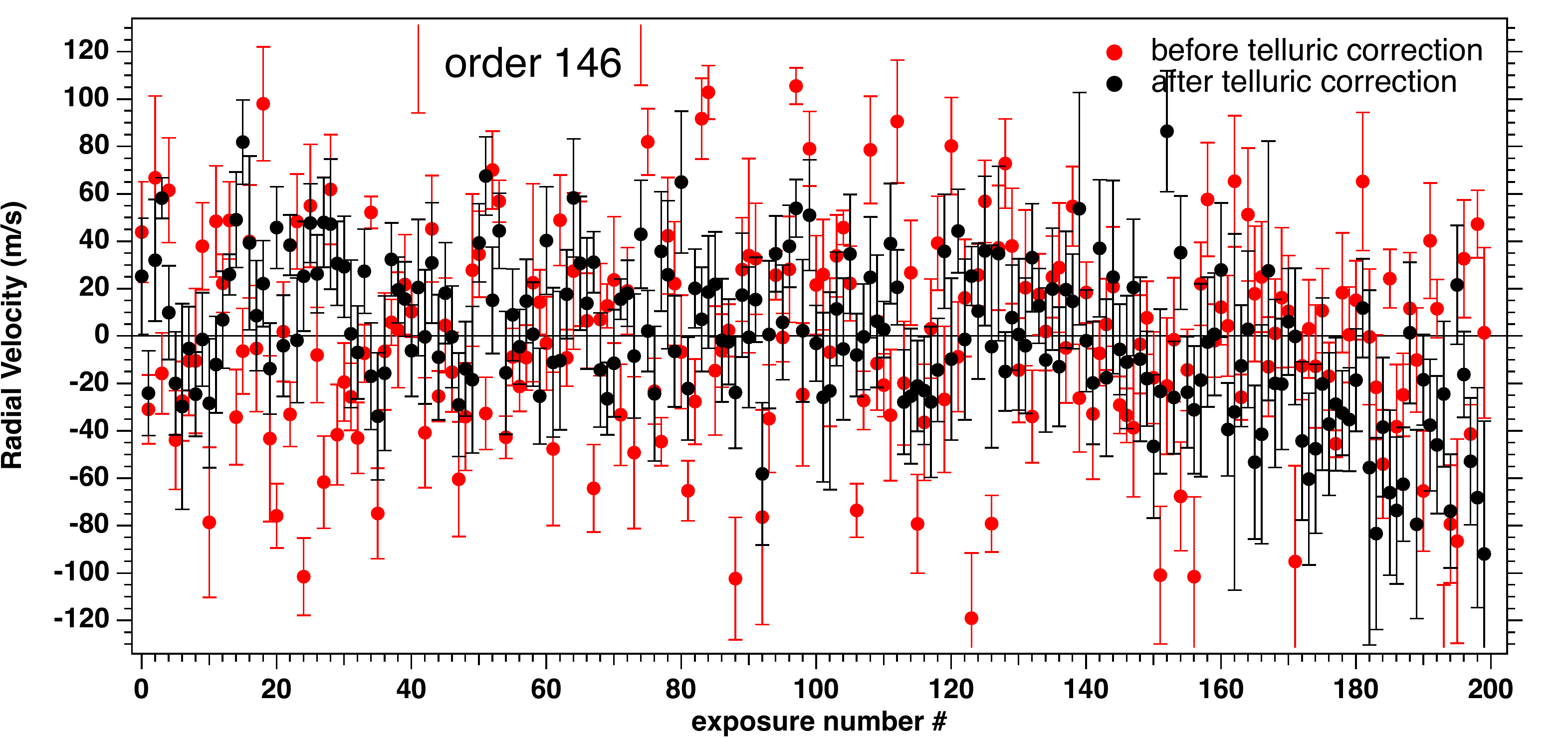}
\caption{Radial velocity of star HD40307 for the extreme case of order 146, which is heavily contaminated by O$_{2}$. In red - result with the standard binary mask BM1 before telluric correction; in black – result with our new binary mask BM2 after telluric correction. The average radial velocities of 31386.4 and 31388 ms$^{-1}$ were subtracted respectively from the series before and after correction.}
\label{figure_13_comparisonorder146}
\end{figure*}

\begin{figure*}
\centering
\includegraphics[width=0.9\hsize]{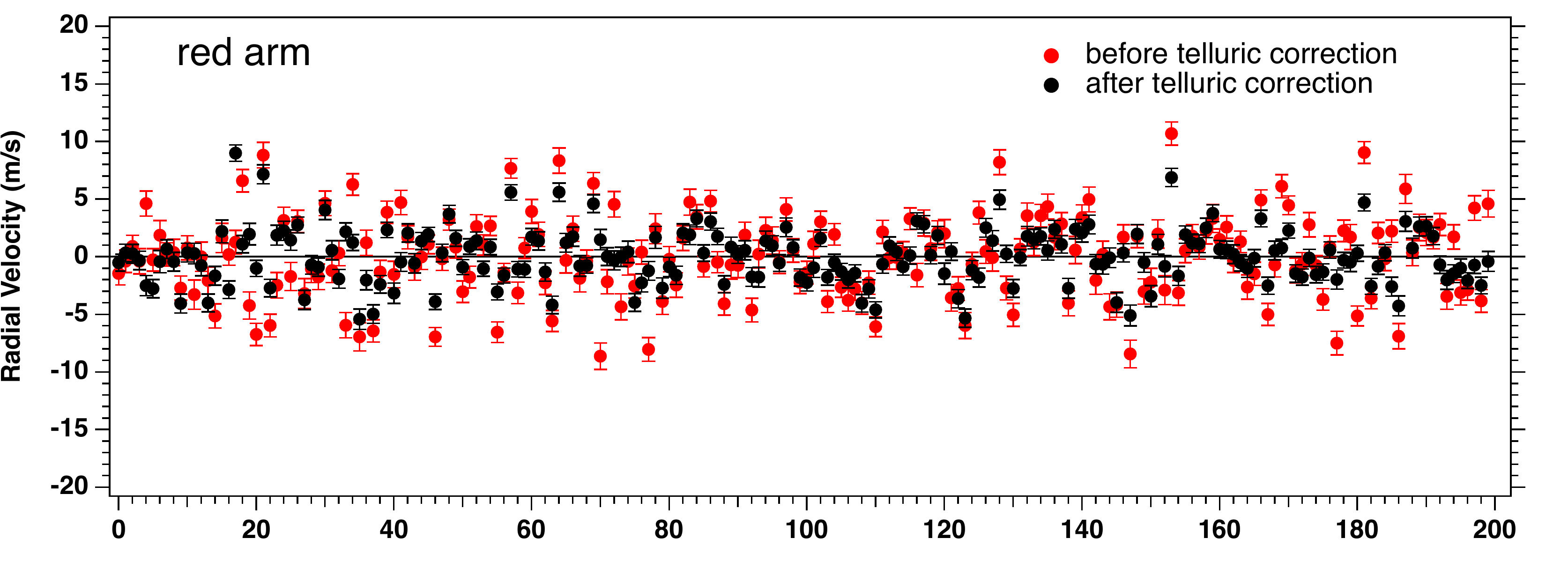}
\caption{Radial velocity of star HD 40307 for orders from 90 to 163. In red - result with standard binary mask BM1 before telluric correction; in black – result with our new binary mask BM2 after telluric correction. The average radial velocities of 31382.5 and 31384.4 ms$^{-1}$ were subtracted respectively from the series before and after correction.
}
\label{figure_14_comparisonredarm}
\end{figure*}

\begin{figure*}
\centering
\includegraphics[width=0.9\hsize]{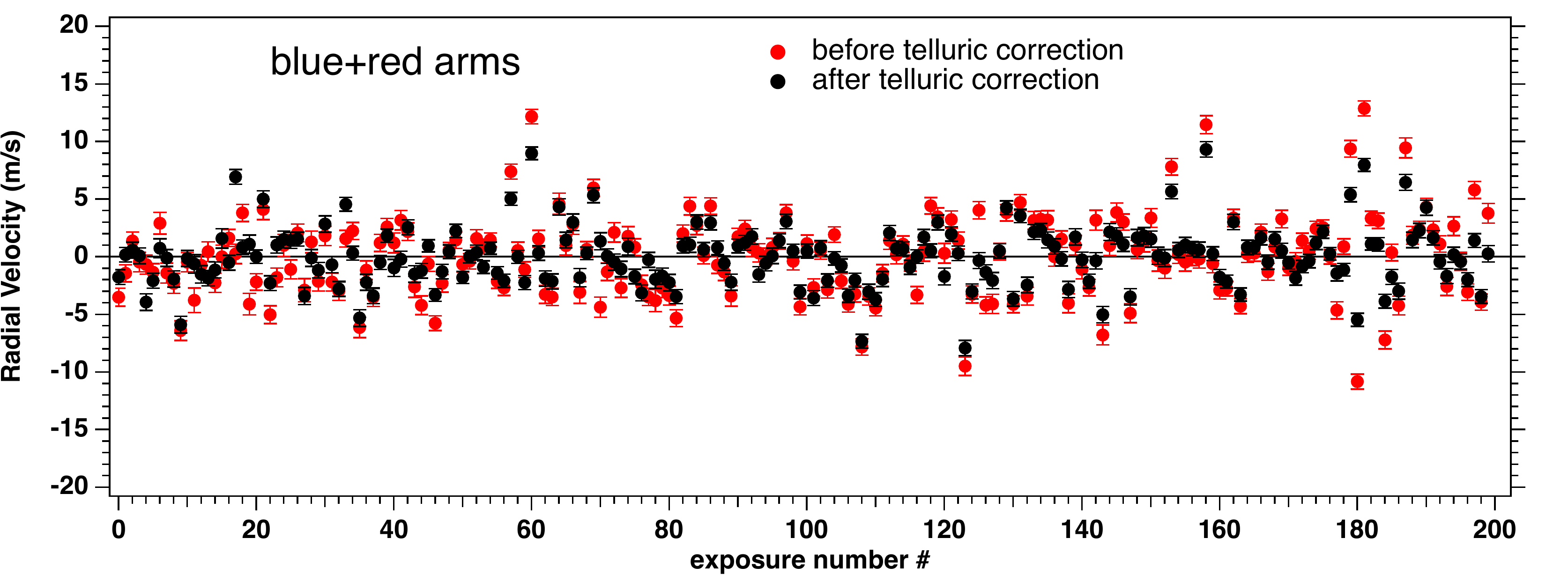}
\caption{Radial velocity of star HD 40307 for orders from 40 to 163. In red - result with standard binary mask BM1 before telluric correction; in black – result with our new binary mask BM2 after telluric correction. The average radial velocities of 31386.9 and 31385.9 ms$^{-1}$ were subtracted respectively from the series before and after correction.}
\label{figure_15_comparisonredblue}
\end{figure*}

 \section{Results}
 \label{results}

 In this section we show a series of comparisons between two separate applications of the CCF code. On one hand, the code is applied to the uncorrected data on the basis of the standard binary mask BM1. On the other hand, it is applied to corrected data on the basis of the combination of the standard mask and the newly created mask, giving BM2. We show three different cases, relatively easy H$_{2}$O correction (116-117), moderately easy O$_{2}$ correction ($\gamma$ band, orders 130-131), and the more difficult case of the O$_{2}$ B band (orders 146-147). 

 \subsection{Order 130 contaminated by O$_{2}$}  
 
% We have corrected all 200 exposures of HD 40307. After we created a new binary mask based on the standard one and corrected observed spectra.
The order 130 ($\lambda$ $\simeq$6316 \AA~) and its twin order 131  are contaminated by the $\gamma$ band of O$_{2}$ (figure \ref{figure_8_order_130_mask}), and only 3 lines are contained in the official BM1 of ESPRESSO, while our new mask BM2, as described above, contains 38 lines for this order.

In figure \ref{figure_11_comparisonorder130} are compared the RV measurements (for the whole series of 200 exposures) coming from our CCF algorithm, before correction, and after correction of tellurics. The average (over time) of RV has been subtracted from the whole series, to ease the comparison of the dispersion. In principle, we should find a constant signal, over the $\simeq$ 4 hours of measurements.  We see by eye that the signal is about constant in average. Both the error bars and the spread of measurements (fluctuations, quantified by the standard deviation S$_{dev}$) are very much decreased with the telluric corrected data. Quantitatively, the average error bar  of this order over the 200 measurements is reduced from 22.5 to 5.3 ms$^{-1}$ for the new mask, while the spread of RV measurements S$_{dev}$ is reduced from 35.4 to 7.0 m s$^{-1}$. This is of course a somewhat extreme example of what can be gained by the addition of new portions of star spectra, once they are corrected from tellurics. Independently of this improvement, the constancy of the RV value over the exposures constitutes also a test of the telluric correction. 

\subsection{Order 116 contaminated by H$_{2}$O} 
Similarly, the order 116 ($\lambda$ $\simeq$ 5891 \AA~) and its twin order 117  are contaminated by the blue end of a region contaminated by H$_{2}$O lines (figure \ref{figure_1_mask}). BM1 contains only 5 lines, while BM2 contains 20 lines for this order.

In figure \ref{figure_12_comparisonorder116} are compared the RV measurements, before correction, and after correction of tellurics. Again, the time averaged RV has been subtracted. The average error bar of this order is reduced from 12.8 to 7.1 ms$^{-1}$ for the new mask, while the spread of RV measurements S$_{dev}$ is reduced from 20.5 to 9.6 m s$^{-1}$. The vertical scale of RV covers the same extent of 260 ms$^{-1}$.

Figure \ref{figure_13_comparisonorder146} is identical to Figure \ref{figure_11_comparisonorder130}, but for order 146, which is heavily contaminated by the strong O$_2$ B band (see Figure \ref{figure_10_order_146_mask}). We include this order as an extreme case since a number of telluric lines are saturated before convolution by the PSF and there are strong residuals. As we already mentioned, a better correction would require a specific determination of the exact shape of the instrumental PSF, which is beyond the scope of this work. The standard binary mask contains only 3 stellar lines, while our binary mask created after telluric correction contains 18 stellar lines. Here, the spread decreased from 47 ms$^{-1}$ to 32 ms$^{-1}$. However, the formal error increased from the rather high value of 21 to 25 ms$^{-1}$, a somewhat puzzling result. We note a similar behavior for order 144, which is not contaminated by telluric absorption.

\subsection{Red arm}
 
 We compare in Figure \ref{figure_14_comparisonredarm} the series of RV measurements when the orders of the red arm are combined together, with a weight 1/err$^{2}$ where err is the formal error returned for each order. Note that the red arm contains the orders 90 to 169, while we kept only orders 90 to 163, dropping the four orders 164-167 heavily contaminated by the strongest A band from O$_{2}$, and containing saturated lines. Moreover, the last orders 168-169 are characterized by a low signal. The vertical scale of RV covers a 40 ms$^{-1}$ full extent. 

The average error bar of this red arm is reduced from 1.04 to 0.78 ms$^{-1}$ for the new mask BM2 (Table \ref{Table_results}). If we use the photon-noise estimation on the CCF (see Appendix \ref{appendix}), the error-bar decreases from 0.89 to 0.72 ms$^{-1}$, a reduction factor of 1.24. As discussed in the Appendix, the difference is due to our choice of using 10 points of the CCF instead of the CCF in its entirety. Therefore, it is likely that the photon-noise estimator is more robust than the 10 points Gaussian fit. For both cases, this gain is entirely due to additional photons from the 696 new stellar lines from BMc included in the spectral regions made available by the telluric correction. The spread of RV measurements S$_{dev}$ is reduced from 2.83 to 2.37 ms$^{-1}$ (see Table \ref{Table_results}).  The reduction of the spread, a factor 1.19, is slightly smaller than the reduction of the formal error, 1.33.

\subsection{Red+Blue arms}

Finally, we performed the same comparison as above for a combination of red and blue arm data. More precisely, we combined data for orders 40 to 163 coming from both arms. Orders below 40 were dropped because of a very low signal for this K2.5 star. The comparison is shown in Fig. \ref{figure_15_comparisonredblue}. The average error bar of blue+red arms is reduced from 0.77 to 0.64 ms$^{-1}$ for the new mask BM2, while the spread of RV measurements S$_{dev}$ is reduced from 3.53 to 2.59 ms$^{-1}$ (see Table \ref{Table_results}). The vertical scale of RV covers again a 40 ms$^{-1}$ full extent. The reduction of the spread, a factor 1.36, is slightly larger than the reduction of the formal error, 1.20. 

\subsection{Absence of RV trend in RV time series}

\begin{figure*}
\centering
\includegraphics[width=0.9\hsize]{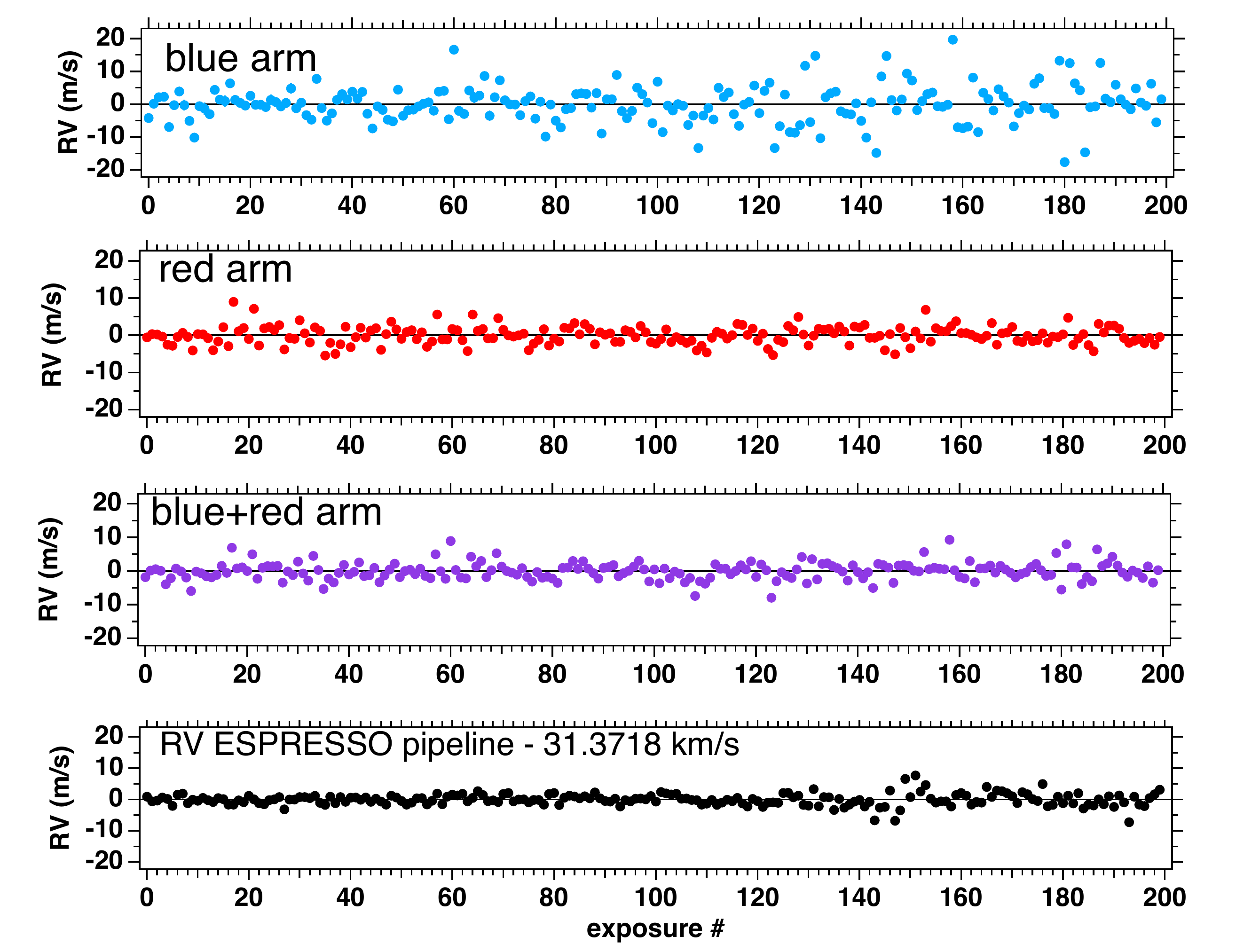}
\caption{Temporal and BLUE-RED dependence of the measured RV. Top blue graph: RV measurements from our processing after telluric correction restricted to blue orders. Top red graph: Same for the red orders. Bottom purple graph: combination of blue and red orders above.
Bottom black graph: RV measurements reported by \citep{Pepe2021}.}
\label{figure_16_jitter_redblue}
\end{figure*}

On figure \ref{figure_16_jitter_redblue} we compare four time series of RV measurements:
- the original RV measurements found in the archive (similar to Figure 21 of \cite{Pepe2021}. 
- the RV measurements from our processing after telluric correction restricted to blue orders (from 40 to 89, $\lambda$ from 4319 to 5206 \AA~ ). In fact, it is almost identical to our processing of data not corrected from tellurics (not shown here).
- our RV measurements after telluric correction restricted to red orders (from 90 to 163, $\lambda$ from 5195 to 7504 \AA~)
- the RV measurements after telluric correction for combination of blue and red orders.
All three curves have been obtained after subtraction of their mean value, and are plotted on the same scale. We recall that we did not do any filtering and used a threshold for the contrast, and for this reason we do not obtain the same final result for each RV point than \cite{Pepe2021}. All the time series show a constant value, with some fluctuations that will be discussed further in Section \ref{comparison}. A linear fit to the red arm RV values indicates a change of RV (a drift) of less than  0.064 m/s (the error bar encompasses 0), while the BERV had a change of 240 m/s over the whole series: the telluric system of lines has moved w.r.t. the star system of lines by this amount along the time series. The fact that there is no measurable trend over the time series is a serious indication that the telluric system has been indeed well removed in the telluric corrected spectra. Unfortunately, the time series that we have analyzed covers only 240 m/s, while the BERV could change by $\simeq$ 30 km/s, corresponding to a maximum drift of 8 m/s. The campaign of HD40307 observations covered only a limited excursion of BERV and we cannot do much better with this campaign. However, \cite{Allart22} have indeed shown on Tau Ceti that possible drifts due to uncorrected remnants of tellurics are smaller than $\simeq$  0.2 m/s over a large BERV excursion ($\pm$28 km s$^{-1}$, their figure 12) in the case of their correction technique. Since we do an accurate computation of the atmospheric transmission (by taking the whole p-T vertical profile into account instead of a mono-layer) and are using the same HITRAN molecular database, we may expect that our atmospheric correction scheme would not produce more drift. The study of the exact amount of drift is an interesting subject, but beyond the scope of this paper. It would require another target, since the planetary orbits are not yet known with enough precision to detect drifts on the order of 0.2 m/s.

\begin{figure*}
\centering
\includegraphics[width=0.9\hsize]{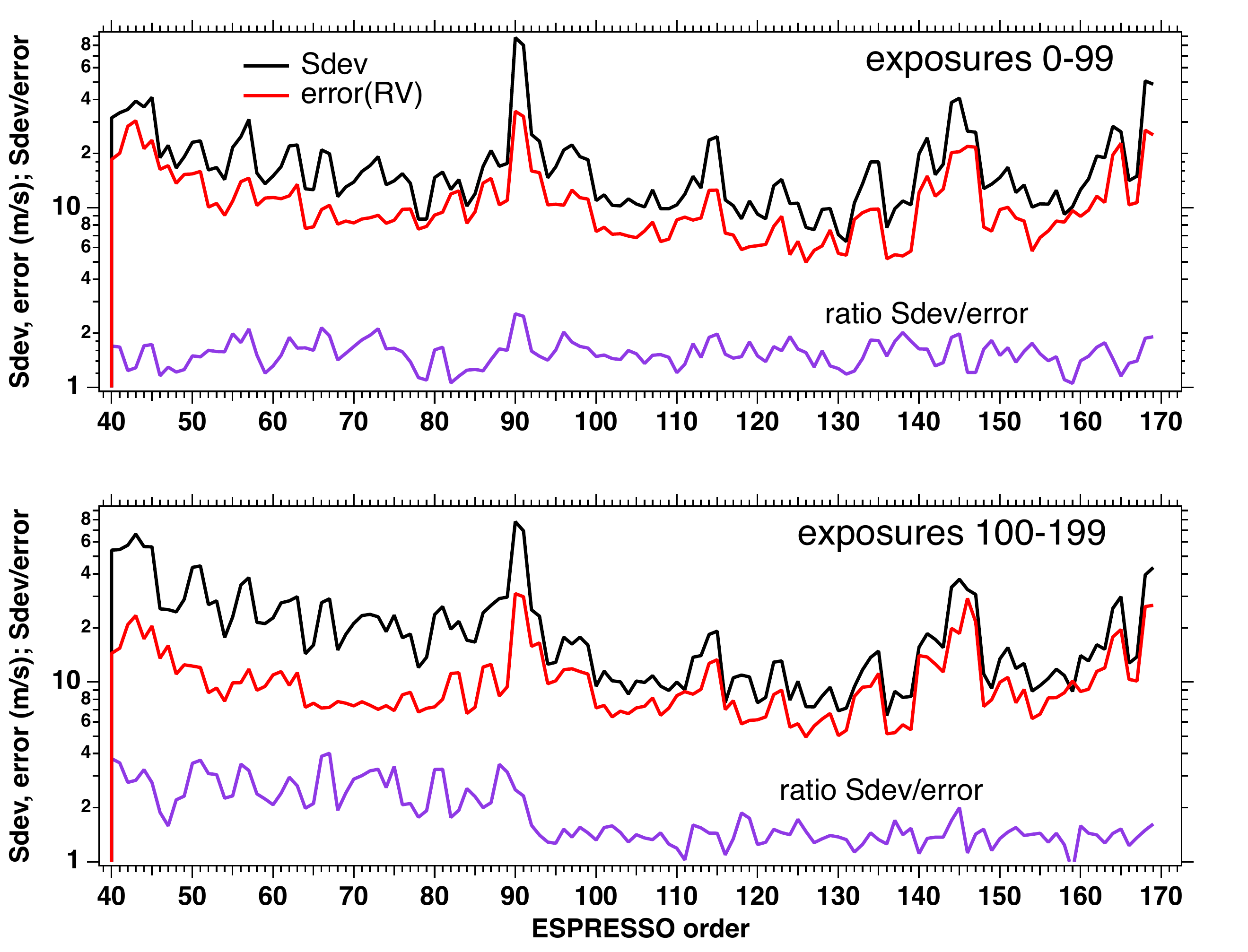}
\caption{Wavelength and temporal dependence of the observed RV. Top: Standard deviation S$_{dev}$ (black line) and formal error on RV measurements (red line) for each order from our processing, averaged on the first 100 exposures only. The ratio between the two quantities is shown in violet. Bottom: Same as top for the last 100  exposures. Note the high ratio found in the blue orders during the second half of the observational time , a sign of spurious jitter due to a blue cryostat thermal instability.}
\label{figure_17_jitterSdev}
\end{figure*}

\section{Temporal and wavelength dependence of radial velocity fluctuations}
\label{comparison}
During our analysis we found out the presence of RV signal fluctuations, as previously mentioned in \citet{Pepe2021}.  What we call the jitter here are such rapid fluctuations of the RV signal, with amplitudes much above the formal error, without prejudging the physical nature of the cause of this jitter. The existence of the time series of 200 exposures of star HD40307 offers the opportunity to identify some time periods where the dispersion of RV values are larger than usual (Figure \ref{figure_16_jitter_redblue}). In order to investigate further about these fluctuations, we considered separately the blue and red arms of ESPRESSO data.

\subsection{Temporal dependence}

It is striking to see the difference in observed jitter between the blue arm and the red arm. While the blue arm reproduces the time variation of the jitter presented in \cite{Pepe2021}, with a sudden increase approximately after exposure 130, the red arm does not see such an increase.
We divided the time series into two groups: first group for exposures 0-99, and second group for exposures 100-199. The change of the standard deviation S$_{dev}$ from group 1 to group 2 is quite different for the blue arm and the red arm, as seen in Table \ref{Table_results}. Quantitatively, going from group 1 to group 2, for the blue arm, S$_{dev}$ increased substantially from 4.13 to 6.71 ms$^{-1}$. At the same time, for the red arm,  S$_{dev}$ decreased (slightly) from 2.50 to 2.33 ms$^{-1}$.

%Quantitatively, the standard deviation S$_{dev}$ of the red arm, computed on exposures 100-199, is only 2.23 m s$^{-1}$  while it is three times larger for the blue arm, at 6.71 m s$^{-1}$ (Table \ref{Table_1}) 

Our results in Figure \ref{figure_16_jitter_redblue} imply that the fluctuations occur only on the blue detector. It is a little worrisome to see this dichotomy arriving between the blue arm and the red arm, since they have two different detectors, and since the telluric correction was mostly relevant to the red arm. In any case, such a change in radial velocity only detected by only one arm of the spectrograph can hardly be related to stellar p-mode oscillations, because implied Doppler shifts of the stellar atmosphere would affect the RVs equally the red and the blue part of the spectrum. Fortunately, our referee pointed out to us that until a major instrument intervention in May 2022, ESPRESSO had a blue cryostat thermal stability problem \citep{Figueira21}. Therefore this type of fluctuations (jitter) affecting only the blue arm that we have detected is certainly caused mainly by the thermal instability of the blue detector, and not related to the onset of p-mode oscillations. 
%Thanks to a comment from our referee, we learned that until a major instrument intervention in May 2022, ESPRESSO had a blue cryostat thermal stability problem, a problem we were not aware of. Therefore this type of fluctuations was very likely caused by thermal instability of detector.
The fact that our separate blue arm and red arm measurements clearly revealed this behavior shows that our telluric correction, mostly applied in the red arm, does not introduce artefacts which would mask this difference in red and arm fluctuation levels.

\begin{table*}
\caption{CCF statistical results. Units are m/s.}             % title of Table
\label{Table_results}      % is used to refer this table in the text
\centering                          % used for centering table
\begin{tabular}{c c c c c}        % centered columns (4 columns)
\hline\hline                 % inserts double horizontal lines
 Used exposures & 0-199  & 0-199 & 0-99 & 100-199 \\    % table heading 
                 & no correction  & with correction & with correction & with correction \\
\hline                        % inserts single horizontal line
   Formal error err & 1.04 & 0.78 & 0.79 & 0.78\\    
     Red Arm &  &  &  & \\
    Formal error err  & 1.15 & 1.15 & 1.21 & 1.10 \\    Blue Arm &  &  &  & \\
      Formal error err  & 0.77 & 0.64 & 0.65 & 0.63 \\    Red + Blue Arms &  &  &  & \\
Standard deviation S$_{dev}$ & 2.83 & 2.37   & 2.50 &2.23\\
Red Arm  &  &     &  &\\
Standard deviation S$_{dev}$ & 5.78 & 5.56  & 4.13 & 6.71\\
Blue Arm  &  &     &  &\\
Standard deviation S$_{dev}$ & 3.53 & 2.59 & 3.03 & 3.97\\
Red + Blue Arms  &  &     &  &\\
%Ratio S$_{dev}$/error & 2.72 &  3.03   & 3.16 & 2.86\\
%Red  Arm  &  &     &  &\\
%Ratio S$_{dev}$/error &  &     &  & \\
%Blue  Arm  & 5.02 &  4.83   & 3.41 & 6.1\\
\hline                                   %inserts single line
\end{tabular}
\end{table*}
%
%     Ratio err/S$_{dev}$ & &     &  &\\

\subsection{Wavelength dependence}

In order to refine the wavelength dependence of the jitter (does it co\"{i}ncides with the blue arm only?),  the time series of retrieved RVs were also studied for each order separately. In particular, we examined the variation versus the ESPRESSO order
of three variables: the mean (formal) error err of RV averaged over
exposures, the standard deviation S$_{dev}$ of the RV values among
exposures, and the ratio S$_{dev}$/ error, displayed on Figure \ref{figure_17_jitterSdev} as a function of order number (equivalent to wavelength) and separately for the two groups of exposures. The idea behind is that, in the frame of the theory of Gaussian errors, the standard deviation S$_{dev}$  should be equal to the mean error of a series of measurements. This would hold if the error were well estimated, and if the spectra were not changing (otherwise from shot noise), implying a constant retrieved RV (within shot noise limits).

We note on figure \ref{figure_17_jitterSdev} that both the error and S$_{dev}$ are changing, for three reasons: the intensity of the spectrum in the order (shot noise), the number of mask lines for this order and their depths (contrast). 

However, the value of S$_{dev}$ is systematically larger than the value of the error, with a strong correlation (from order to order) between S$_{dev}$  and the error. There are usually two reasons for which the standard deviation is found larger than the mean error: either the error bar has been underestimated, or there are time fluctuations (jitter) in the data superimposed to a constant mean RV value, or a combination of both effects.

It can be shown mathematically ( with the theorem of the variance of a sum of independent variables X and Y) that S$_{dev}$ may be related to the formal error and the jitter amplitude by: 
\begin{equation}
    S_{dev}^2= jitt^2+k^2err^2    
\label{eqA1}
\end{equation}

where  {\it jitt} is the amplitude of the jitter (affecting the RV), and  k is a factor by which the formal error is underestimated w.r.t. the true error. This factor  k is likely valid at all times and all orders, because it is intrinsic to the retrieval algorithm.

For the second group of exposures (100-199), a dichotomy is observed between the left part (blue arm) and the right part (red part). The behaviour of the right part is quite similar to the whole top curve (0-99), both for the correlation between error(RV) and S$_{dev}$ , and the about constant value of the ratio around 1.41 (actually
lower than for the first group at 1.55 for the red arm). On the contrary, for the blue arm, the S$_{dev}$  curve is detached above the RV curve, there is no correlation between error(RV) and S$_{dev}$, and the ratio is much larger, with a mean of 2.67. Therefore, the jitter which appears mostly during the second half of the series of exposures, happens only in all the orders of the blue arm.

\begin{figure}
\centering
\includegraphics[width=0.8\hsize]{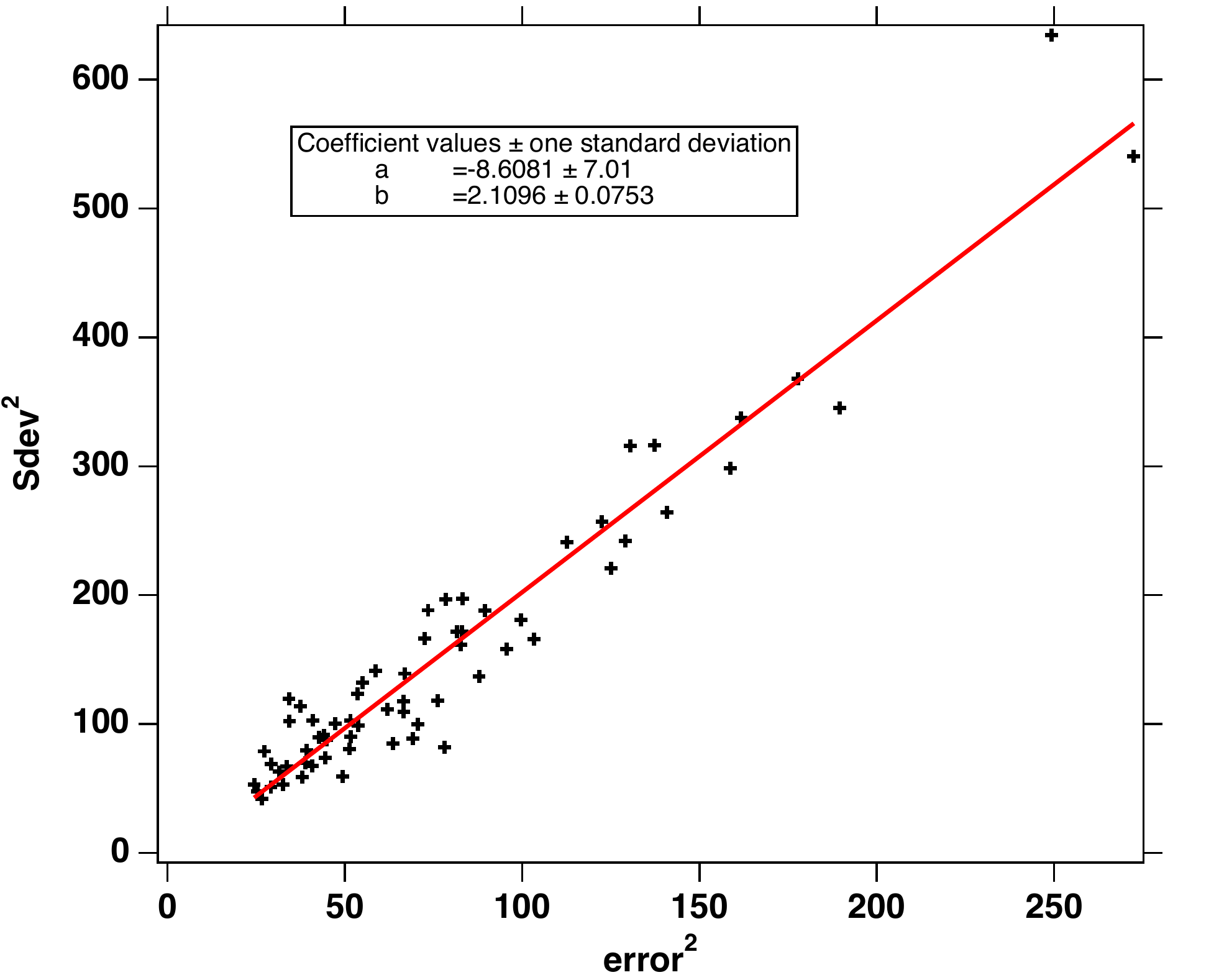}
\caption{Linear fit of the scatter plot of S$_{dev}^{2}$ as a function of the error$^2$. Each point is for one red order of ESPRESSO. The slope allows to determine the factor k of underestimation of formal error: k= true error/formal error.}
\label{figure18}
\end{figure}

As said above, the explanation of this blue jitter is instrumental, and linked to thermal instabilities of the cryostat containing the blue detector. The smallest values of the ratio S$_{dev}$/err are found in the red arm, for exposures 100-199 (figure \ref{figure_17_jitterSdev} ). Let
us assume that the jitter amplitude is the same for all red orders
in this time frame. On figure \ref{figure18} is plotted S$_{dev}$$^{2}$ as a function of
the squared formal error err$^{2}$, each point for one order from 92 to
169. After the exclusion of a few outliers, there is a clear linear
relation ship of the form a + bx , where the slope b = k$^{2}$  is found to be 2. 1 $\pm$  0. 07. This means that our formal error is underestimated (w.r.t. the true error) by  a factor $\simeq$ 1.4. 

Extrapolating the linear fit to err = 0 the value of
S$_{dev}^{2}$  becomes equal to the jitt$^{2}$. The value is found equal to -8.6 $\pm$ 7 m$^{2}$s$^{-2}$.  Of course a squared quantity cannot be negative, and this means that the jitter is likely weak, undetectable in this portion of spectrum and exposures 100-199.

\section{Conclusions}
\label{conclusion}
We have used an exceptional series of consecutive short exposures of the ESPRESSO spectrograph for a K2.5V type star to test and estimate the potential increase in precision of the RV measurement brought by telluric line removal when this correction is based on a publicly available, location- and time-matched atmospheric transmission spectrum and a simple and conventional fitting code to be performed locally. \\
We used theoretical transmissions of H$_2$O and O$_2$ downloaded from the TAPAS on-line facility and adjusted them to each exposure, then divided each recorded spectrum by this adjusted transmission. TAPAS conducts a top-quality computation of the telluric absorption with the meteorological field giving the vertical T-P profile, HITRAN spectroscopic data base and LBLRTM code. To perform the adjustment of H$_{2}$O and O$_{2}$ columns, we took benefit from the characteristics of the target star, a temperature of $\simeq$ 4980K and a low metallicity of $\simeq$ -0.3, resulting in lines narrow and weak enough to allow easy stellar continuum fitting and disentangling between tellurics and stellar features. This has made possible to mask the stellar lines, adjust the telluric transmissions in a very precise way and divide the data by the adjusted telluric model, without a direct use of a stellar synthetic spectrum.
The quality of the adjustment is illustrated by the good correspondence between the fitted H$_2$O column and the column deduced for each exposure from the infra-red radiometer located at Paranal, as well as the good correspondence between the adjusted O$_2$ column and the airmass.\\ 
We augmented the ESPRESSO binary mask with 696 additional stellar lines located within regions previously excluded due to the presence of tellurics (additional mask BMc). We applied the same standard CCF method, with the same parameters to initial data and standard binary mask on the one hand, and to corrected data and the augmented mask on the other hand. \\
We have compared the average error on RV as well as the average spread of the results before and after the telluric correction and mask extension. For data from orders 90 to 163 (the ESPRESSO red arm) we found that the average error on RV measurements decreased from E(BM1)= 1.04 m$s^{-1}$ for the standard binary mask BM1 to E(BM2)= 0.78 m$s^{-1}$ when using spectra corrected from telluric absorption and the new binary mask adapted to corrected regions. The blue arm is practically not affected by the telluric absorptions. 
The red arm error values of E(BM1)=1.04 and E(BM2)=0.78 m $s^{-1}$1  are for a 30 s exposure of the star HD40307, spectral type K2.5, visual and red magnitudes V= 7.1, R=6.6, giving  $\simeq$ 7094 electrons per pixel at center of order 146 (SNR$\simeq$84). For the same stellar spectral type, we expect that both errors will scale inversely to the SNR per pixel (from shot noise), therefore their ratio will remain constant. For cooler stars which have more flux in the red part for the same visual magnitude V, the reduction factor of the error will be larger.
On the other hand, in the course of our study of the series of measurements of HD40307, we noted that some jitter affected the RV signal, revealing some sudden increase of stellar activity, as reported earlier \citet{Pepe2021}. We found out that this jitter surprisingly affected much more the bluest part of the spectrum than the red part. We recently learned that this fluctuation was produced by thermal fluctuations of the blue arm cryostat.

When using the full spectrum (blue+red arms), the formal error after telluric correction is reduced from 0.77 to 0.64 m s$^{-1}$ (or, if using the photon noise minimal uncertainty estimation, from 0.55 to 0.51 m s$^{-1}$ ). This improvement corresponds to a saving of about 45\% of telescope time to reach the same precision (or 16\% of telescope time if using the photon noise minimal uncertainty estimation). This is important, when considering the huge need for telescope time for the monitoring of targets to determine their population of planets. Furthermore, it has been shown that long-term stellar activity (e.g., associated to the rotation period of the star) induces some RV spurious variations that are smaller in the red than in the blue, at least in the case of Proxima Centauri  \citep{Suarez2020}. This increases the interest of telluric corrections, since they broaden the spectral intervals available for RV retrievals. Similarly, all stellar spectra taken before the correction of the blue detector instability (introducing a short-term artificial “jitter” in the RV data derived from the blue detector) could be reprocessed with our TAPAS derived procedure for a better RV estimate.

%\textcolor{red}{In the course of our study of the series of measurements of HD40307, we noted that some jitter affected the RV signal, revealing some sudden increase of stellar activity, as reported earlier \citep{Pepe2021}. However, we found out that this jitter affected much more the bluest part of the spectrum than the red part (whose jitter actually decreased a little while the blue jitter increased much). We have developed some arguments to show that this finding is somewhat consistent with the correlation between the shift of individual Fe lines of the solar spectrum with their EWs, recently measured by \cite{Gonzales2020}. Studying the time series of exposures of HD40307 allowed documenting the sporadic presence of jitter. However, the usual strategy of exoplanets survey programs is to take only one exposure per star, and looking at many stars during one night. In such a case, it is difficult to detect the presence of jitter. But since the jitter is found to be much smaller in the red, it increases substantially the interest of telluric correction which adds new spectral window in this wavelength region.} 

 The binary mask BMc built for the telluric contaminated regions is probably valid also for other spectral types not to far from K2.5 but this need to be explored. Cooler-type stars with more numerous lines (e.g., M stars) and/or stars with higher metallicity might  be more difficult to treat, however state-of-the-art atmospheric transmission spectra and synthetic stellar spectra allow to predict atmospheric lines or fractions of lines free of stellar absorption. Moreover, it is has been shown that in this case, fitting the observed spectrum with a forward model of the product of a stellar synthetic spectrum and a telluric transmission spectrum is also feasible \citep[see, e.g.][]{Puspitarini15}, allowing determining the quantity of H$_{2}$O and verify the quantity of O$_{2}$  in the atmosphere at the time of observations. Such procedures allow to retrieve the stellar spectrum corrected from atmospheric absorption.

The present technique has the advantage of being applicable to any individual recorded spectrum, i.e it does not require time series or libraries of spectra. It benefits from the very highly detailed transmission spectra publicly available from the TAPAS facility. Telluric lines associated with observatories located at different altitudes and in different regions have different shapes, and these differences are not entirely smoothed out after convolution by the spectrograph PSF. There are also changes according to the season. This is why taking advantage of the computations done for a stratified atmosphere adapted to the location and the date is providing an optimal input for the corrections. As we showed, the O$_{2}$ transmission spectrum can be directly used after scaling to the airmass, while the H$_{2}$O transmission spectrum requires a scaling. As a matter of factor, the atmospheric water vapor is changing rapidly, often on timescales smaller than an hour, and the 6 hours interval between ECMWF computations is too long to follow its variations. We have shown that the H$_{2}$O columns we have adjusted for all individual exposures follow very closely the values derived from the Paranal infrared radiometer and indicated in each exposure data file. This implies that, if an on site measurement of precipitable water vapor column is provided, an initial scaling can be performed using the ratio between this measured column and the one predicted by TAPAS/ECMWF, indicated in the accompanying information. Equivalently, one may enter the on site value in the TAPAS request form and the computed profile will be adjusted accordingly. This provides a preliminary H$_{2}$O transmission very close to the true solution.

The corrections we have presented here require to download TAPAS transmission spectra. TAPAS computations are now very fast and interruptions are very rare since the website has been upgraded. Because species others than H$_{2}$O follow the airmass, as we have shown here for O$_{2}$, it is possible to retrieve TAPAS transmissions for the different species at the beginning of the observing night or during an exposure and to adapt them to the airmass of each observation. Adjustments of H$_{2}$O are required for each exposure, as said above, but they are also very fast provided one has prepared in advance the mask appropriate to the stellar type. For this work, the H$_{2}$O adjustment, the computation of the profiles, order by order, and the correction take less than one minute by exposure on a laptop with a 2.4 GHz Intel Core I9. TAPAS now covers the 300-3500 nm interval and is well adapted to both the near UV, e.g. for the CUBES spectrograph under development \citep{Covino22}, and to NIR spectrographs such like SPIROU \citep{Donati20} or NIRPS \citep{Bouchy19}. The available species in TAPAS are O$_{2}$, H$_{2}$O , O$_{3}$, CO$_{2}$, CH$_{4}$ and N$_{2}$O. NO$_{2}$ and NO$_{3}$ are also expected to be available soon.

\begin{acknowledgements}

We thank our referee for his/her very constructive comments. We deeply thank Piercarlo Bonifacio for providing his personal ATLAS-SYNTHE environment and helping us to compute an appropriate stellar synthetic spectrum, used for comparison tests (template matching) and stellar line identification. We acknowledge a very useful support from Burkhard Wolff from ESO who helped us with ESPRESSO instrumental details and the binary mask. We thank the anonymous referee for his useful comments and for having pointed out the instrumental reason of increased jitter for the blue detector data.
A.I. acknowledges the support of a Vernadski Scholarship for PhD students, sponsored by the French Government and the Ministry of Science and Higher Education of the Russian Federation under the grant 075-15-2020-780 (N13.1902.21.0039).
This research has made use of the SIMBAD database, operated at CDS, Strasbourg, France.
\end{acknowledgements}

\bibliographystyle{aa}
\bibliography{mybib.bib}

\appendix
\begin{appendix}
\section{The RV uncertainty estimation based on photon-noise and CCF.}
\label{appendix}
\begin{figure*}
\centering
\includegraphics[width=0.9\hsize]{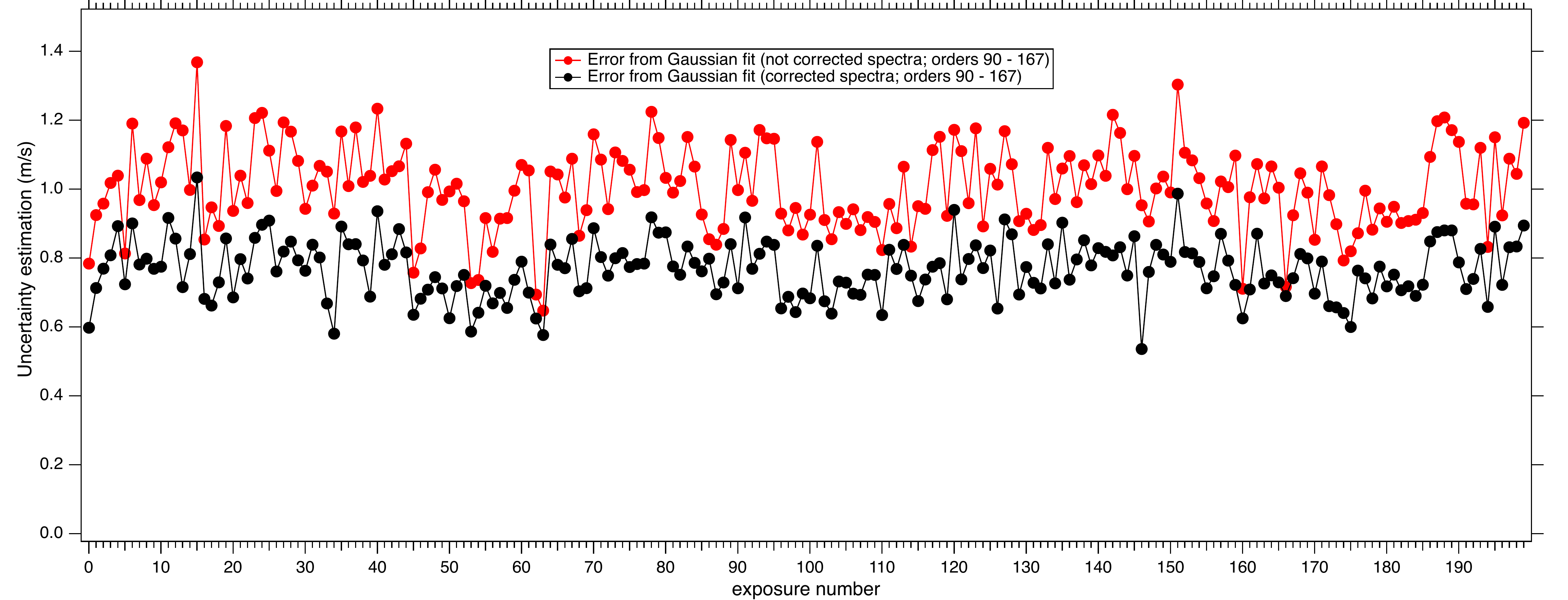}
\caption{Estimated uncertainty $\delta V_{G}$ based on the Gaussian fit for the red arm. Red curve - before telluric correction; black curve - after telluric correction}
\label{figureappendix1}
\end{figure*}

\begin{figure*}
\centering
\includegraphics[width=0.9\hsize]{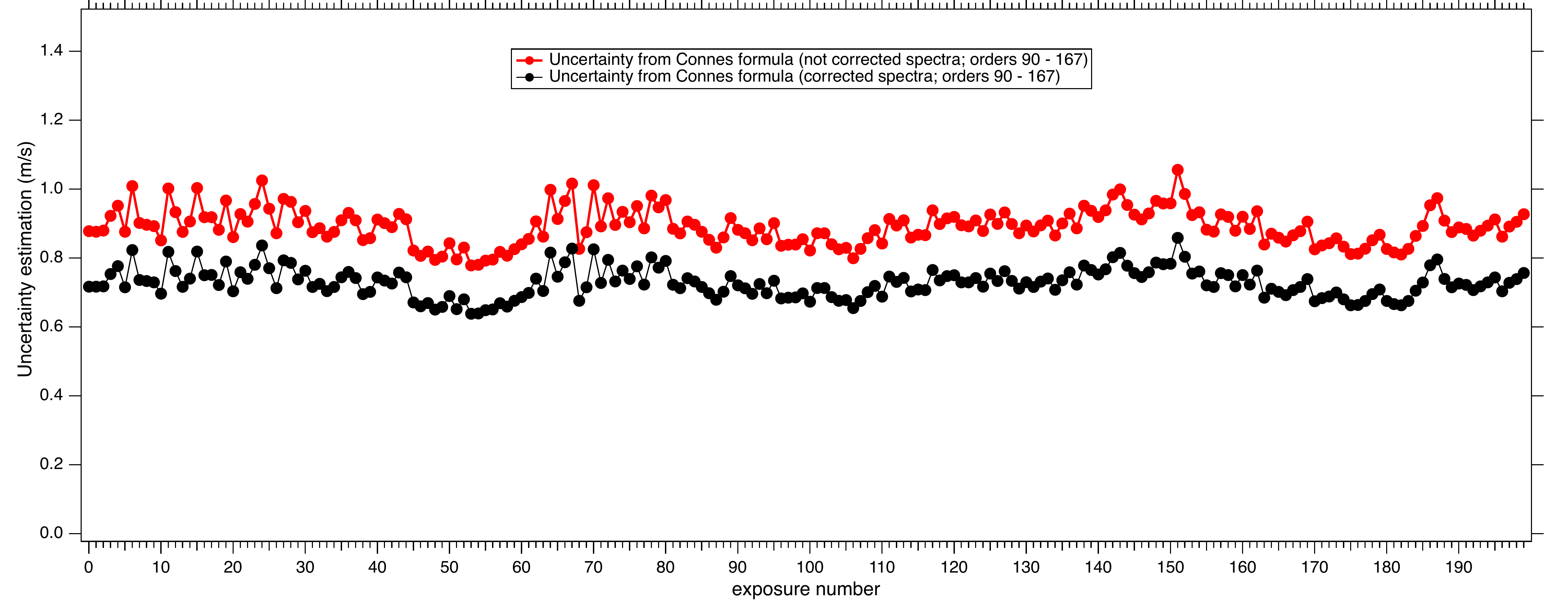}
\caption{Estimated uncertainty $\delta V_{QCCF}$ based on the quality factor for CCF from \cite{Boisse2010} for the red arm. Red curve - before telluric correction; black curve - after telluric correction. Fluctuations are connected to the fluctuations of spectral intensity (see text).}
\label{figureappendix2}
\end{figure*}

\begin{figure*}
\centering
\includegraphics[width=0.9\hsize]{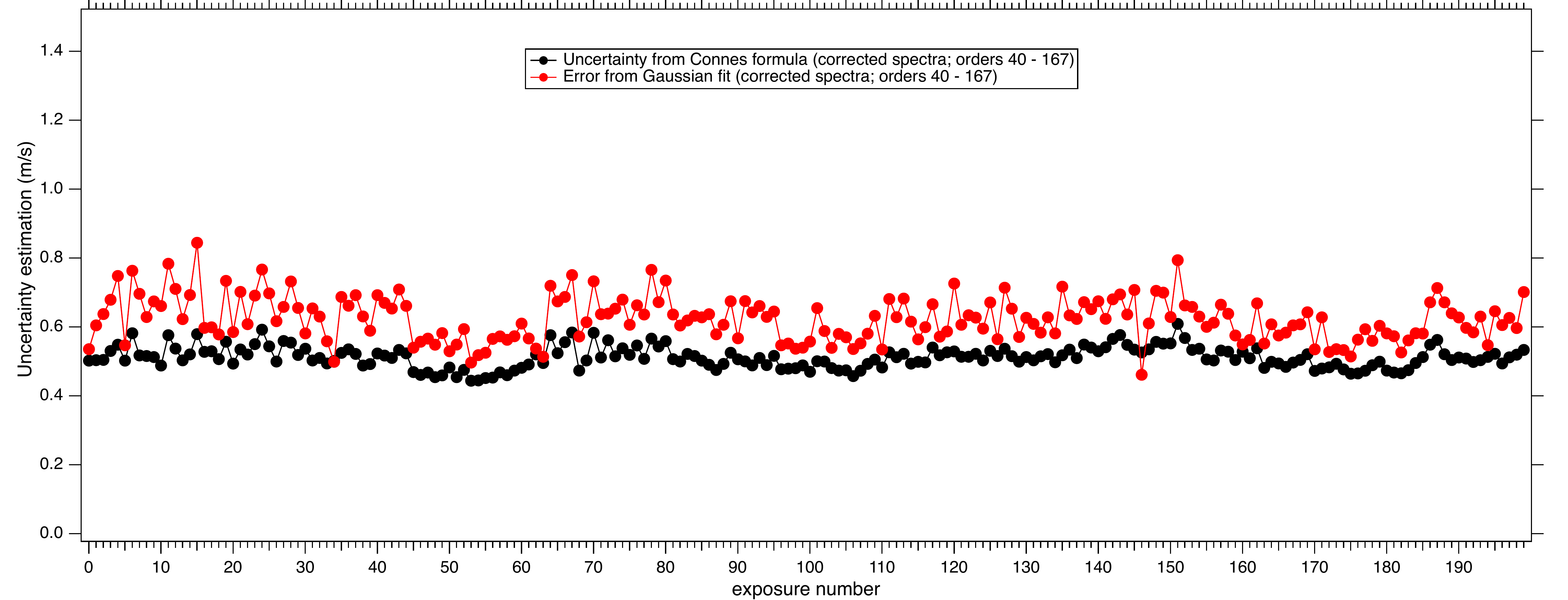}
\caption{Comparison of estimated uncertainties: based on Gaussian fit of the CCF ($\delta V_{G}$, in red) and  on the quality factor for the CCF from \cite{Boisse2010} ($\delta V_{QCCF}$, in black), for red and blue arms combined.}
\label{figureappendix3}
\end{figure*}

The photon noise is one limiting factor for the precision of radial velocities (RV) time-change measurements, the other being the richness of the spectral features to be used to detect a Doppler shift of the star. A mathematical expression of the limitation was developed by \cite{Connes1985} and tested for different types of stars and spectrographs by \cite{Bouchy2001}. It uses the quality factor Q devised by \cite{Connes1985} and revived in \cite{Bouchy2001}, which represents the spectral line richness of the spectrum; it is independent of the actual size of spectral range, but function of the spectral profile "wigginess" within that range. The estimation of the radial velocity uncertainty $\delta V_{Q}$ based on the quality factor Q is: 
\begin{equation}
\delta V_{Q}= \frac{c}{Q\sqrt{N_{e^-}}}
\label{eqApp1}
\end{equation}
where $c$ – speed of light, Q – quality factor, $N_{e^-}$ - the total number of photo-electrons counted over the whole spectral range. The quality factor Q makes use of the derivative of the intensity with respect to the wavelength and should be estimated from a noise-free spectrum. The derivative is computed by finite difference of intensity from one spectel to the next.\\
In reality an observed spectrum is not noise free. Therefore, with noisy data the quality factor will be spuriously higher and the RV uncertainty will be lower than the true optimal uncertainty.\\
In \cite{Boisse2010}, the authors described an approach to apply this formalism, not to a spectrum as did \cite{Connes1985}, but to a CCF. Since the CCF for an order is constructed by piling up all lines of this order it can be considered as an average spectral line of the order and also as noise-free. In \cite{Boisse2010} the quality factor of the spectrum is calculated as:
\begin{equation}
Q=\frac{\sqrt{\sum W\left(i\right)}}{\sqrt{\sum{A_0\left(i\right)}}}
\label{eqApp2}
\end{equation}
where i is the pixel number of the spectrum,  $W\left(i\right)=\frac{\lambda^2\left(i\right)\left(\partial A_0\left(i\right)/\partial\lambda\left(i\right)\right)^2}{A_0\left(i\right)+\sigma_D^2}$ , with $A_0$ – reference noise free spectrum intensity and $\sigma_D^2$ – detector noise.\\
In the case of the CCF, the wavelength scale is replaced by  a velocity scale. The quality factor $Q_{CCF}$ shown in \cite{Boisse2010} is becoming:\\
\begin{equation}
Q_{CCF}=\frac{\sqrt{\sum_{i}{\left(\frac{\partial CCF\left(i\right)}{\partial V\left(i\right)}\right)^2/CCF_{noise}^2\left(i\right)}}}{\sqrt{\sum_{i} CCF\left(i\right)}}\sqrt{N_{scale}}
\label{eqApp3}
\end{equation}
where $CCF_{noise}(i)$ is the quadratic sum of the photon noise and the read-out detector noise integrated inside the CCF mask holes for the velocity i – analogous to the $A_0(i)+\sigma_D^2$  in W(i) formulation from \cite{Bouchy2001}; $N_{scale}$ – correction factor corresponding to the scale of the velocity step in detector pixel unit. The velocity uncertainty is equal to:
\begin{equation}
\delta V_{QCCF}=\frac{1}{Q_{CCF}\sqrt{\sum CCF\left(i\right)}}
\label{eqApp4}
\end{equation}
Uncertainties computed both from our Gaussian fit of the CCF $\delta V_{G}$ and from the application of the above quality factor to the CCF $\delta V_{QCCF}$ are shown on figures \ref{figureappendix1} and \ref{figureappendix2}. In both cases estimated uncertainties decrease after telluric correction. We have verified that the fluctuations of $\delta V_{QCCF}$ (Fig. \ref{figureappendix2}) are almost entirely  due to intensity fluctuations reflected on $\sqrt{N_{e-}}$. This explains the parallelism of both curves, before and after telluric correction. On the last figure \ref{figureappendix3} we show a comparison between our estimates based on a Gaussian fit $\delta V_{G}$ and those based on the quality factor $\delta V_{QCCF}$ for all orders (blue and red) simultaneously. We can see that generally the uncertainty estimation obtained from the quality factor $\delta V_{QCCF}$ is smaller than the one obtained by Gaussian fit and also displays smaller fluctuations. This is because \cite{Boisse2010} used the full CCF to perform their estimate while we used only 10 points of the CCF for our Gaussian fit (we recall that in this way we attempt to avoid RV biases appearing when fitting a  whole spectral line as described by \cite{Gonzales2020}.

\end{appendix}
\end{document}